\documentstyle[aps,epsfig,preprint]{revtex}
\draft
\tighten
\begin{document}

\title{Investigation of the high momentum component of 
        nuclear wave function using hard quasielastic A(p,2p)X reactions.}
\author{ I.~Yaron, J.~Alster, L.~Frankfurt, E.~Piasetzky}
\address{School of Physics and Astronomy,  Sackler Faculty of Exact Sciences, 
Tel Aviv University, Ramat Aviv 69978, Israel}
\author{M.~Sargsian}
\address{Department of Physics, Florida International University, Miami, FL 33199, U.S.A}
\author{M.~Strikman}
\address{Department of Physics, Pennsylvania State University, University Park,
PA 16802, U.S.A}
\maketitle
\begin{center}
\today
\end{center}

\begin{abstract}
We present theoretical analysis of the first data on the
high energy and momentum transfer (hard)  quasielastic $C(p,2p)X$
reactions. The cross section of hard $A(p,2p)X$  reaction is calculated
within the light-cone impulse approximation based on two-nucleon
correlation  model for the  high-momentum component of  the nuclear wave
function. The nuclear effects due to  modification of the  bound nucleon 
structure, soft nucleon-nucleon reinteraction in the initial and final 
states of the reaction with and without color coherence have been considered.
The calculations including these nuclear effects show that the distribution 
of the bound proton light-cone momentum fraction $(\alpha)$ shifts towards
small values ($\alpha < 1$), effect which was previously derived only 
within plane wave impulse approximation. 
This shift is very sensitive to the strength of the
short range correlations in nuclei. Also calculated is an excess of the
total longitudinal momentum of outgoing protons.
The calculations are  compared with data on the $C(p,2p)X$
reaction obtained from the EVA/AGS experiment at Brookhaven National
Laboratory.
These data show $\alpha$-shift in agreement with the calculations.
The comparison allows also to single out the contribution
from short-range nucleon correlations. The obtained strength of
the correlations is in agreement  with the values previously obtained
from electroproduction reactions on nuclei.
\end{abstract}

\newpage

\section{Introduction}
\label{intro}

One of the important signatures of quark-gluon structure in nucleon-nucleon 
interaction at short distances is the observed strong energy dependence 
($\sim s^{-10}$) 
of the wide angle pp elastic differential cross section at $s\geq 12 ~GeV^2$,
where $s$ is the square of the pp c.m.  energy.
Despite the ongoing debate on the validity of perturbative QCD in this
energy region \cite{hex,Isgur_Smith,Rady} or the debate on the
relevance of a particular mechanism of subnucleon interaction (i.e. 
quark-interchange\cite{BCL79,FGST79,RS95}, three-gluon 
exchange\cite{LLP,BoSt}, reggeon-type contribution\cite{BoSo}), it is commonly 
accepted that the power-law $s$- dependence of the elastic cross 
section signals the  onset of the  hard dynamics of the quark-gluon interaction.

In this paper we address the question of what happens when wide angle pp 
scattering takes place inside the nucleus, i.e. the incident proton is scattered 
off a bound proton. If this reaction would have the 
same $\sim s^{-10}$ energy dependence as that of the cross section of free
$pp$ scattering, one may expect that the incoming proton 
will favor to  scatter off  a bound proton with larger initial momentum 
aligned to the direction of  the incoming  proton \cite{FS88,FLFS}.  
This kinematic condition  corresponds to   $pp$ scattering with smaller $s$
and therefore  larger scattering cross section. Thus, if 
nuclear effects will not alter the genuine $s$-dependence of the $pp$ 
cross section, the high momentum transfer $p+A\to p + p + X$ reaction 
would select preferably  the high momentum components of the 
nuclear wave function.  

Due to the short-range nature of the strong interaction, the  high 
internal momentum in the nucleus will be generated mainly by 
short-range NN correlations. Therefore, at sufficiently  high energies 
and high momentum transfers one expects to probe the short-range 
properties of the nucleus. 
In Ref.\cite{FS88,FLFS}  within a plane wave impulse approximation~(PWIA) 
the authors calculated the cross section of high momentum transfer
$A(p,2p)X$ reactions and observed  a  strong sensitivity to
the high momentum component of the nuclear wave function.

Motivated by the recent measurements of high-momentum transfer pA 
reactions at Brookhaven National Laboratory (BNL)\cite{kn:I101}
we carried out  a detailed analysis of the high-momentum transfer 
$A(p,2p)X$  reaction  investigating specifically the competing nuclear 
effects,  not discussed  previously. These effects may obscure the
observed  sensitivity  shown  within PWIA~\cite{FLFS}.
Our main goal is to see whether  these reactions probe short range 
correlations (SRC) and their sensitivity to the dynamical 
structure of these correlations.

The structure of the paper is as follows.
In Chapter 2 we outline the basic theoretical framework for the 
calculation of  the high-energy wide angle quasielastic $A(p,2p)X$ reaction.
We also discuss the  nuclear effects which can compete 
with the expected signatures of the   scattering from SRC.
In Chapter 3 we present  the predictions of the model presented in
the Chapter 2. Chapter 4 describes briefly the EVA experiment at BNL.
The calculations are compared with the data obtained in this 
experiment in chapter 5. In chapter 6 we summarize the results of our
study. 

\section{The Basic Theoretical Framework}
\label{II}

 In quasi-elastic (QE) scattering a projectile is elastically scattered
from a single bound ``target'' nucleon in the nucleus while the rest of
the nucleus acts as a spectator. A schematic presentation of (p,2p) QE
scattering is given in Fig 1. 

\begin{figure}[ht]
  \begin{center}
    \leavevmode
    \centerline{\epsfig{file=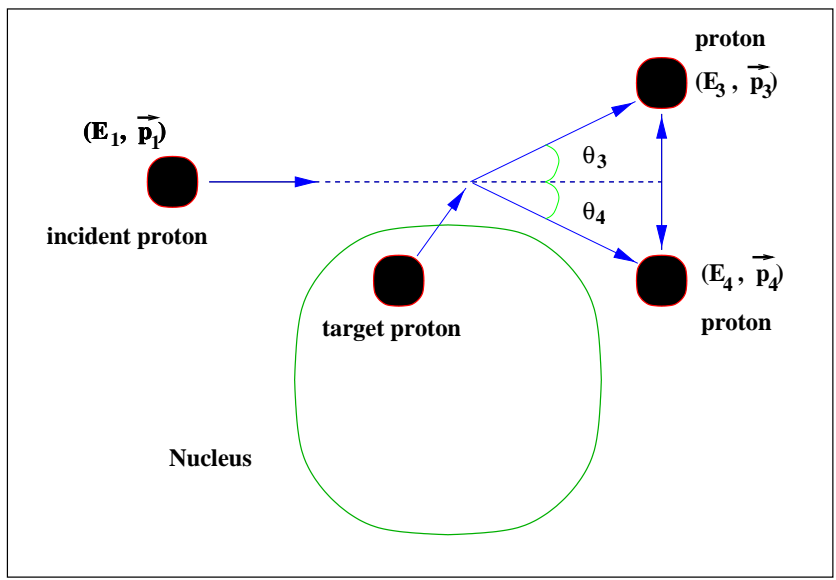,width=11.4cm,height=8.cm}}
   \caption{The kinematics for quasi-elastic $A(p,2p)X$ scattering.}
  \end{center}
\label{Fig.1}  
\end{figure}

\subsection{\bf Kinematics }
\label{IIa}

${\em p_{A}} =(E_{A},\vec{p_{A}})$, ${\em p_{1}}=(E_{1},\vec{p_{1}})$,
${\em p_{3}}=(E_{3},\vec{p_{3}})$, ${\em p_{4}} =(E_{4},\vec{p_{4}})$,
${\em p_{R}} =(E_{R},\vec{p_{R}})$    - are the four - momenta of the target
nucleus, the incoming proton, the scattered proton, the ejected proton and
the recoil nucleus, respectively. For simplicity we did not show in the
Figure~1 ${\em p_{A}}$ and ${\em p_{R}}$.
Using the variables defined in the figure the Mandelstam variables are:

\begin{eqnarray}
s & = &  ({\em p_3}+{\em p_4})^{2} \ ; \ t = ({\it p_1}-{\it p_3})^{2}.
\label{kinst}
\end{eqnarray}

\vspace{0.4cm}

The high-momentum transfer primary process in the $A(p,2p)X$ quasi-elastic 
reaction is the hard $pp$ elastic scattering.
Since the general predictions are based on the implication of  the strong 
$s$-dependence  ($\sim 1/s^{10}$) of hard elastic $pp$ cross section we will 
limit our calculations to high energy and high momentum transfer 
kinematics  were the $1/s^{10}$-dependence is observed experimentally 
for $pp$ scattering off hydrogen target. Thus, our calculations are  limited
to  $s\gtrsim 12 \ GeV^2$ and $\theta_{cm}\sim 90^0$.

The missing energy ($E_m$) for the $A(p,2p)X$ reaction is given by 
$E_m$ = $E_{1}$ + $E_A$ $-$ $E_3$  $-$ $E_4$ $-$ $E_{A-1}$.
The available high energy $A(p,2p)X$ data have a missing energy 
resolution of about 240 Mev \cite{kn:I101}. 
Therefore, the calculations which we compare with the data are integrated  
over a wide range of missing energy. This integration simplifies the 
calculations as will be  discussed below.

\subsection{Plane Wave Impulse Approximation}
\label{IIb}

A clear interpretation of the quasi-elastic measurements is possible in 
Plane Wave Impulse Approximation (PWIA). Within this approximation 
it is possible to separate nuclear properties from the reaction mechanism. 

In a high energy scattering the reaction evolves near the 
light cone $\tau = t-z\sim 1/(E+p_z)\ll t+z$, where z is the direction of 
the incident proton and $E$, $p_z$ are the 
energy and leading longitudinal momentum of the high energy particles 
involved in the scattering. Thus it is natural to describe the reaction 
in the light cone reference frame (similar to high energy deep-inelastic
scattering from  hydrogen  target see e.g. \cite{Feynman}). 

Within the light cone plane wave impulse approximation the cross section 
of the quasielastic $A(p,2p)X$ reaction can be represented as a convolution  of the
elementary elastic $pp$ scattering cross section off bound nucleon and 
the four - dimensional Light Cone Spectral function\cite{FS88}:
\begin{eqnarray}
\frac{d^{6}\sigma}{(d^{3}p_{3}/2E_{3})(d^{3}p_{4}/2E_{4}) } &   & =  
\sum\limits_{Z}{1\over 4j_{pA}}{{\vert M_{pp} \vert}^{2} \over (2\pi)^2}\cdot 
{P_{A}(\alpha,p_{t}^{2},p_{R+}) \over \alpha^{2}} = \nonumber \\
&   &                                                \nonumber \\
&   & = \sum\limits_{Z}{2\over \pi}\sqrt{s^{2}-4m^{2}s} 
\frac{d\sigma}{dt}^{pp}(s,t)
\cdot {P_{A}(\alpha,p_{t}^{2},p_{R+})\over A\cdot\alpha}
\label{pwia}
\end{eqnarray}
where 
\begin{eqnarray}
p_2 & = & p_3 + p_4 - p_1 \ ; \ p_{t} =  p_{3}^t + p_{4}^t \nonumber \\
\alpha & = & \alpha_{4} + \alpha_{3} - \alpha_{1} \ ; \ 
\alpha_{i} =   A{p_{i-}\over P_{A-}} \equiv  A{E_{i}-p_{i}^z \over 
E_{A}-P_{A}^z}. 
\label{kinvar}
\end{eqnarray} 

The superscript $"t"$ and $"z"$ denote the transverse ($x,y$)
and longitudinal directions with respect to incoming proton momentum 
$\vec p_{1}$. 
The $"+"$ and $"-"$ indices denote the energy and longitudinal components of 
four - momenta in the light cone reference frame
\footnote{Since $z$ directions is chosen as the direction of incoming 
proton momentum, the ``-'' component corresponds to the 
light cone longitudinal momentum, which is conserved 
at the scattering vertices.}. 
The variable $\alpha$ defined in Eq.(\ref{kinvar}) describes the light cone 
momentum fraction of nucleus carried out by target nucleon, normalized in 
such a way that a nucleon at rest has $\alpha=1$.
The $j_{pA}$ - is the invariant  flux with respect to the nucleus, 
the $M_{pp}$ and $ \frac{d\sigma}{dt}^{pp}$ - are the invariant amplitude and 
cross section for  elastic  $pp$ scattering.

The Light Cone spectral function represents the  probability to find the 
target nucleon with the light cone momenta ($\alpha$, $p_{t}$) 
times the probability that  the residual nuclear system has a  
momentum component $p_{R+} = E_{R} + p_{R}^{z}$. The Spectral 
function is normalized as follows\cite{FS88}:  
\begin{equation}
\int {p_{A-}\over 2A}P_{A}(\alpha,p_{t}^{2},p_{R+}){d\alpha\over \alpha}
d^{2}p_{t}dp_{R+} = A.
\end{equation} 

\medskip
\medskip

\subsection{\bf The Light Cone Spectral Function}
\label{IIc}

The integration over a wide range of the missing energy 
allows us to use the  following  approximations for 
the spectral function:

For target proton momenta below the Fermi sea level 
($p_{2}< p_{Ferm}\sim 250 MeV/c$ ) we use the nonrelativistic limit of the light 
cone spectral function\cite{FS81,FS88}:

\begin{equation}
P_{A}(\alpha,p_{t}^{2},p_{R+}) \approx {1\over 2}n(p_{2})\cdot
\delta(p_{R+}-(\sqrt{M_{A-1}^{2}+p_{2}^{2}}-p_{2}^{z})),
\label{mf}
\end{equation}
where $\alpha \approx 1 - p_{2}^z/m$ and 
$\vec p_{2} = \vec p_3 + \vec p_4-\vec p_1$ 
are the missing momentum components of the reaction. $n(p)$ is the 
momentum distribution of nucleons calculated within the mean field 
approximation.

For the momentum range of  ($p_{Ferm} < p_{2} < 0.7~ GeV/c )$ we assume  
the dominance of the two nucleon short-range correlations which allows to 
model the  spectral function as follows\cite{FS88,DFSS}:
\begin{eqnarray}
P_{A}(\alpha,p_{t}^{2},p_{R+}) & \approx & \int {A^{2}\over 
2p_{A-}}a_{2}(A)\cdot
\rho_{2}^n\left ({2\alpha\over (A-\beta)},(\vec p_{t}+{\alpha\over 
(A-\beta)}\vec p_{(A-2)t})^{2}\right )
\cdot \rho_{A-2}(\beta,p_{(A-2)t}^{2})\cdot \nonumber 
\\
&   & \nonumber \\
&   & \delta \left (p_{R+}-{m^{2}+(\vec p_{(A-2)t}+\vec p_{t})^{2}\over m 
(A -\alpha-\beta)} -
{M_{A-2}^{2}+p_{(A-2)t}^{2}\over m \beta}\right ){d\beta\over \beta}
d^2p^{t}_{(A-2)},
\label{lcsf2n}
\end{eqnarray}
where ($\beta,p_{(A-2)t}^{2}$) and $\rho_{A-2}$ are the light cone momentum
and the density matrix of the recoiling $(A-2)$ system. The parameter 
$a_{2}(A)$ is the probability of finding  two-nucleon correlations in the 
nucleus A and $\rho_{2}^n$ is the density matrix of the correlated pair 
which we set  equal to  the Light Cone density matrix of the 
deuteron\cite{FS81}:
\begin{equation}
\rho_{2}^n(\alpha,p_{t}^{2}) = {\Psi_{D}^{2}(k)\over 2-\alpha}
\sqrt{m^{2}+k^{2}} \ ; \ 
k = \sqrt{{m^{2}+p_{t}^{2}\over \alpha (2-\alpha)}-m^{2}} \ ; 
\ (0 < \alpha < 2).
\label{hmlc}
\end{equation}
Note that the factorization of the nuclear density matrix to the correlation and 
$(A-2)$ density matrices is specific for the short-range two-nucleon 
correlation approximation. In this approximation it is assumed that the 
singular character of $NN$ potential at short distances (existence of 
repulsive core) defines the main structure of the nucleon momentum distribution 
in SRC and it is less affected by the collective
interaction with the $A-2$ nuclear system.
Notice that the expression in Eq.(\ref{hmlc}) is the light cone analog 
of the approximated spectral function used in Ref.\cite{CSFS}, where 
the validity of two-nucleon correlation approximation is demonstrated 
comparing the prediction of nonrelativistic analogue of  Eq.(\ref{lcsf2n}) 
with the exact calculations of spectral function of $^3He$ nucleus and 
infinite nuclear matter.

To obtain the density matrix of the recoiling $(A-2)$  system, 
additional  physical assumptions are required. However the fact that 
we are interested in the cross section integrated over a wide  range of
the missing energy allows us to simplify the  Eq.(\ref{lcsf2n}) 
by neglecting  
the momentum of the recoiling $(A-2)$ system (SRC at rest approximation):
\begin{equation}
\rho_{A-2}(\beta,p_{(A-2)t}^{2})=
(A-2)\cdot\delta(A-2-\beta)\cdot \delta(p_{(A-2)t}^{2}).
\label{srcrest}
\end{equation}

Inserting Eq.(\ref{srcrest}) into Eq.(\ref{lcsf2n}) one obtains 
the following expression for the light cone spectral function in the 
high missing momentum range:
\begin{equation}
P_{A}(\alpha,p_{t}^{2},p_{R+})  \approx 
{A^{2}\over 2p_{A-}}a_{2}(A)\cdot
\rho_{2}^n(\alpha,p_{t}^{2})\cdot
\delta(p_{R+}-{m^{2}+p_{t}^{2}\over m(2-\alpha)}-M_{A-2}).
\label{sfunrest}
\end{equation}
It is worth noting that the above approximation is justified based on 
the observation of Ref.\cite{CSFS} that it correctly predicts 
the position of the maximum in the missing energy distribution 
at fixed values of missing momenta. 
Therefore, in  regime  in which the integration over the wide range of 
missing energies is allowed,  Eq.(\ref{sfunrest})  represents a valid 
approximation of nuclear spectral function at the  SRC domain.
The same model  was  also  used  to describe the inclusive nucleon and pion 
production in kinematics forbidden for scattering off a free nucleon
\cite{FS88,FS81} and  electroproduction\cite{FS81,DFSS}  reactions from nuclei 
at $x>1$ and $Q^2\geq 1~GeV^2$.

\medskip
\medskip

\subsection{\bf Proton-Proton  Elastic Scattering Cross Section}
\label{IId}

The next quantity which is needed to calculate the quasielastic $A(p,2p)X$ 
cross section in Eq.(\ref{pwia}) is the differential cross section of $pp$ 
elastic  scattering.  For  $s \geq 12 \ GeV^{2}$ we use the phenomenological 
parameterization of the free pp elastic cross section. We
assumed a combination of $s$-parameterization at 
$90^{0}$ presented in Ref.\cite{RP} and $\theta_{c.m.}$-parameterization 
in the form suggested in Ref\cite{SBB}: 
\begin{eqnarray}
 \frac{d\sigma}{dt}^{pp} & = & 45.0 {\mu b\over sr GeV^{2}}\cdot
\left ({10\over s} \right )^{10}\cdot (1-\cos{\theta_{c.m.}})^{-4\gamma}\cdot
\nonumber \\
&   &  \nonumber \\
&   & \left [ 1 + \rho_{1}\sqrt{s\over GeV^{2}}\cdot\cos{\phi(s)} + 
{\rho_{1}^{2}\over 4}{s\over GeV^{2}} \right ]\cdot F(s,\theta_{c.m.})  
\label{pp}
\end{eqnarray}
where $\rho_{1} = 0.08$, $\gamma = 1.6$ and $\phi(s) = {\pi\over 0.06}
\ln(\ln[s/(0.01 GeV^{2})])^{-2}$. The function $F(s,\theta_{c.m.})$ is  
used for further adjustment of  the phenomenologically motivated 
parameterization to the experimental data at $60^{0} \leq \theta_{c.m.} 
\leq 90 ^{0}$\cite{FPSS95}.

\subsection{\bf Calculation of the $\alpha$-dependence of the Cross Section 
in PWIA}
\label{IIe}

The main quantity in which we are interested is the $\alpha$-dependence of 
the $A(p,2p)X$ quasielastic cross section at fixed-large c.m. angles and 
high momentum transfer. The reason of this choice is twofold: first, the 
$\alpha$-dependence naturally expresses the sensitivity  of the 
$A(p,2p)X$ cross section to the high momentum component of the nuclear 
wave function which will be discussed  below; second,
as it will be demonstrated in the Section \ref{IIf2} the $\alpha$ variable
is not sensitive to the soft initial and final state reinteractions 
of energetic protons with target nucleons. Thus its distribution 
will largely reflect the distribution of the nucleon within the SRC 
without substantial modification due to initial and final state interactions.

  In Figure 2 we present the $\alpha$-dependence of the $^{12}C(p,2p)X$ 
cross section calculated for different values of incoming proton momenta. 
The  calculations are  within the PWIA framework described above. 
Here the c.m.  angle of the $pp\to pp$ scattering is restricted to 
$90\pm 5^0$.  The calculation is done for $^{12}C$ target 
using Harmonic Oscillator momentum distribution $n(k)$ in Eq.(\ref{mf}) 
and high momentum tail of the deuteron wave function calculated, 
using the NN Paris potential  in Eq.(\ref{hmlc}), with $a_2(^{12}C)=5$.

\begin{figure}[hbt]  
\vspace{2.0cm}  
\centerline{  
\epsfig{width=4.2in,file=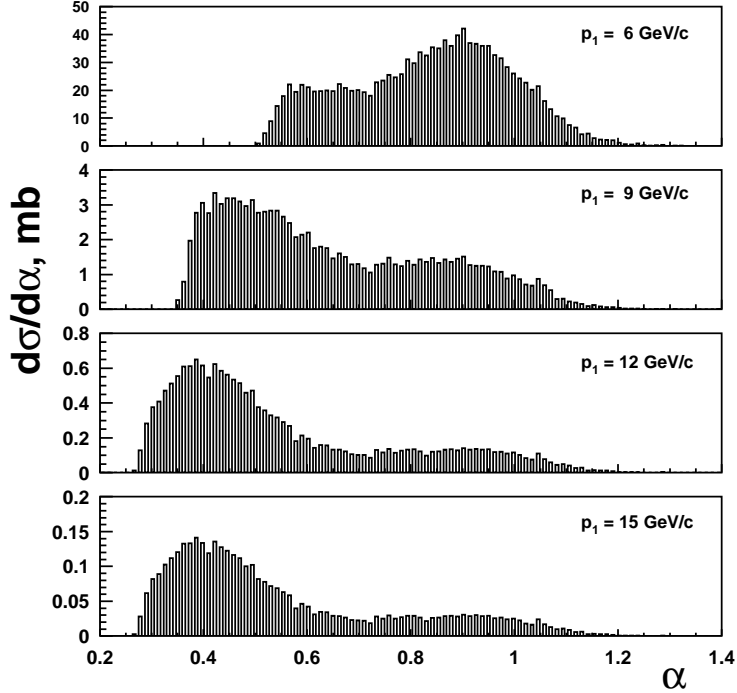}  }
\caption{PWIA calculation of the $\alpha$-dependence of the $^{12}C(p,2p)X$
cross section at different values of incident proton momenta.}  
\label{Fig.2}  
\end{figure}

  Elastic pp scattering off a proton at rest corresponds to $\alpha=1$.
As can be seen from Figure 2, most of the strength is at $\alpha<1$
which corresponds to a scattering off a proton with momenta in the direction 
of $\vec p_1$. This is a quantitative illustration  of the 
discussion in the introduction: the $pp$ cross section on bound
proton scales with the total pp c.m. energy as $\sim (s\alpha)^{-10}$, 
therefore the $A(p,2p)X$ cross section is dominated by smaller $\alpha$.

One can clearly observe a double peak structure of the $\alpha$-distributions.
The first peak, closer to $\alpha=1$, is due to scattering
off a proton in the Fermi sea Eq.(\ref{mf}). The other peak, 
at even lower $\alpha$ values, is due to the scattering off the SRC 
Eq.(\ref{lcsf2n}). As the incoming energy increases, one can see
the shift of the strength to the lower $\alpha$-range which means 
more and more scattering off target protons with high Fermi momenta
aligned in the direction of the incoming proton momentum $p_1$. 
This shift shows the onset of the regime where one expects to 
probe short-range nucleon correlations in the nucleus. 
This picture demonstrates   the selectivity of 
hard $A(p,2p)X$ reactions to the large values of the bound nucleon 
momenta in the nucleus, predicted originally in Ref.\cite{FS88,FLFS}.

\subsection{\bf Competing Nuclear Effects}
\label{IIf}

The calculations above were done within PWIA, using $pp$ 
parameterization (Eq.(\ref{pp})) for the scattering off a free proton. 
Two basic nuclear effects that can obscure the expected $\alpha$-dependence 
are the modification of the bound protons in the nuclei and the initial and 
final state interactions of incoming and scattered protons respectively.

\subsubsection{\bf Nuclear Medium Modification of Bound Protons}
\label{IIf1}

We consider possible binding modifications of  the bound nucleon 
structure which are consistent with the in medium deep-inelastic (DIS) nucleon
structure functions measured using lepton-nucleus  scattering - phenomenon
known as the "EMC effect" \cite{EMC}. One of the  mechanisms that can 
account for the observed modification of DIS structure function   
is the suppression of point-like configurations (PLC)
in a bound nucleon as compared to a free nucleon \cite{FS85,FS88,Frank}.

\begin{figure}[hb]  
\centerline{  
\epsfig{width=4.2in,file=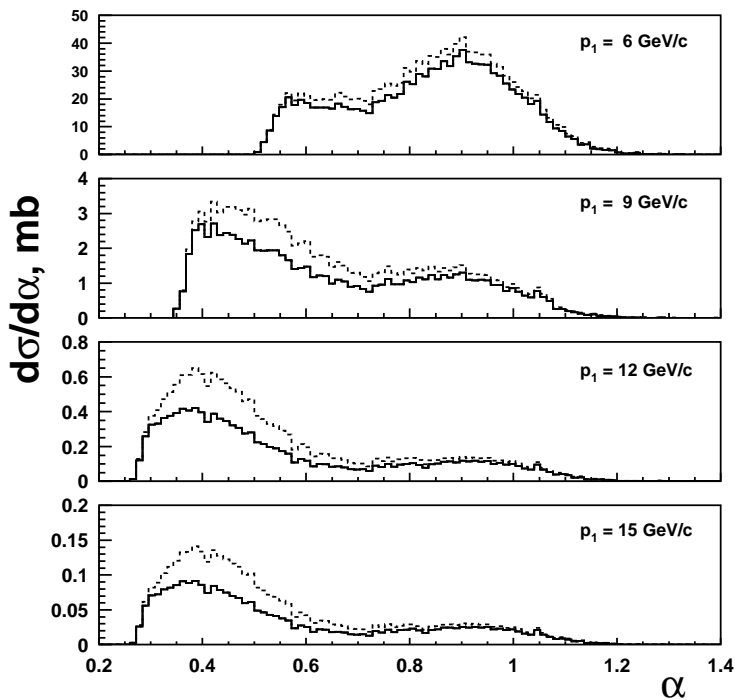}  }
\caption{The $\alpha$-dependence of the cross section for different
values of incident proton momenta. Dashed line - PWIA, solid line- 
with EMC effects discussed in the text. The kinematics are the same as for 
Fig.~2.}  
\label{Fig.3}  
\end{figure}  
 
The PLC are  small sized partonic configurations in the nucleons 
which due to the color screening are weakly interacting objects. 
In the color screening model of EMC\cite{FS85,FS88},
the binding of the nucleonic system results a 
suppression of the nucleon's PLC component. This suppression does not 
lead to a noticeable change in the average characteristics of a nucleon 
in the nucleus. However, it is sufficient to account for the observed EMC 
effect in DIS scattering from nuclei. 
Since the high momentum transfer $pp$ elastic scattering is mainly due to the 
scattering off a PLC in the protons, the expected suppression 
of PLC will reduce the cross section of $pp$ scattering off bound proton. 
This suppression can be estimated by multiplying the 
free $pp$ cross section of Eq.(\ref{pp}) by the factor \cite{FS85}
\begin{equation}
\delta (k,t) = \left( 1 + \Theta (t_{0}-t)\cdot (1- {t_{0}\over t})\cdot
{{k^{2}\over m_{p}} +2\epsilon_{A} \over \Delta E} \right )^{-2},
\label{delta}
\end{equation}
where $\epsilon_{A} \approx 8 \ MeV$ is the average  nuclear binding energy 
and $\Delta E\approx 0.6-1 ~GeV$ is a parameter  that characterize  a typical
excitation of the  bound nucleon. The $t$-dependence in Eq.(\ref{delta}) is 
due to  the fact that in the wave function  of a nucleon the PLC
dominate at sufficiently  high values of the momentum 
transfer \cite{FSZ93} ($-t_0\approx 2 GeV^2$).
As follows from Eq.(\ref{delta}) the $\delta (k,t)$ correction
tends to reduce the expected $\alpha$-shift shown in Figure 1, since it 
introduces additional  $\alpha^l$ (${\l}\sim 2-3$) dependence, 
which softens the  $(s\alpha)^{-10}$ dependence of the $pp$ cross section 
in  Eq.(\ref{pwia}). Note that a similar suppression is expected within 
the rescaling model of the EMC effect \cite{rescaling,MSS}). On the other hand
in a number of models of the EMC effect, such as pion  and binding models 
(for review see \cite{FS88}) the shift to $\alpha<1$ 
is amplified as compared to the multinucleon calculation\cite{FLFS,FMS92}.
Thus, our estimation within  the color screening  model can be considered as 
the upper limit of possible suppression due to binding nucleon modification.

Using  Eqs.(\ref{pwia},\ref{sfunrest},\ref{pp},\ref{delta}) 
the calculated cross section is  shown in Figure 3 as a function of 
$\alpha$. As Figure 3 shows, the considered medium modification effect 
suppresses the high momentum strength of the cross section, 
since it corresponds 
to the larger virtualities of the bound nucleon, which are more sensitive
to the PLC structure of nucleon. However, the suppression does not 
diminish the expected downward shift of the $\alpha$-distribution. It  
would require very unreasonable modifications of the bound nucleon structure 
(contradicting the EMC effects in DIS)  to make the $\alpha$-shift 
(to the $\alpha <1$ region) completely vanish.

\subsubsection{\bf The Effect of the Initial and Final State Interactions}
\label{IIf2}

The major nuclear effect which can obscure the information on SRC 
are the initial and final state interactions (ISI,~FSI) of the incident 
and outgoing protons in the nuclear medium. 
Since the momenta of incoming and two outgoing protons 
are above a few GeV/c one can calculate these rescatterings in 
eikonal approximation. 

For bound nucleons with small momenta $0.8 <\alpha <1.2 $ and 
$p_t\leq p_{Ferm}$, where  the scattered and the knocked-out protons reinteract 
with uncorrelated  nucleons, we apply the conventional Glauber approximation 
to calculate the small angle rescatterings.  This is justified since in 
these cases the spectator nucleons  can be considered as a stationary 
scatterers. 
Integrating  over a wide  range of the missing energy of the $A(p,2p)X$ 
reaction allows to simplify further the calculation of ISI/FSI using 
the probabilistic approximation of  Ref.\cite{Yael}, which
accounts for all orders (single, double, etc) of the soft 
$pN$  rescatterings. 

\medskip

However, the above approximation cannot be used for the bound 
protons in SRC (which have a  large value of Fermi momentum). There, 
the spectator nucleon cannot be treated as a stationary scatterer and 
therefore the Glauber approximation is not valid (see e.g. \cite{FSS}). 
To calculate the initial and final state rescatterings in this case 
we  assume  that for incoming and outgoing protons  
the first rescattering most probably happens with the partner nucleon in 
the SRC. 
Indeed, as it was demonstrated in  Ref.\cite{DFSS}, because of the large 
virtuality of interacting nucleon in SRC the distance of the first soft 
reinteraction after the point of hard interaction is less than $1$~fm 
and it decreases with the increase of $t$ and $p_2$. 
Within the framework of two-nucleon  correlation model one can account for the 
soft rescatterings in the SRC using the calculation of $d(p,2p)n$ reaction   
in Generalized Eikonal Approximation~(GEA), Ref.\cite{FSS,FPSS97}. 
Using the GEA we only calculate the single rescatterings of 
the incoming and  knocked-out protons with the correlated nucleon 
(Figure  4~b,c,d).
The main feature of the GEA is that it accounts for the 
nonzero values of spectator nucleon momentum (it does not treat the
spectator as a stationary scatterer as being done in the conventional
eikonal approximation). This feature is especially
important in the SRC region since in this case the correlated nucleon
momenta are large  and can not be neglected.

\begin{figure}[hb]  
\centerline{  
\epsfig{width=4.2in,file=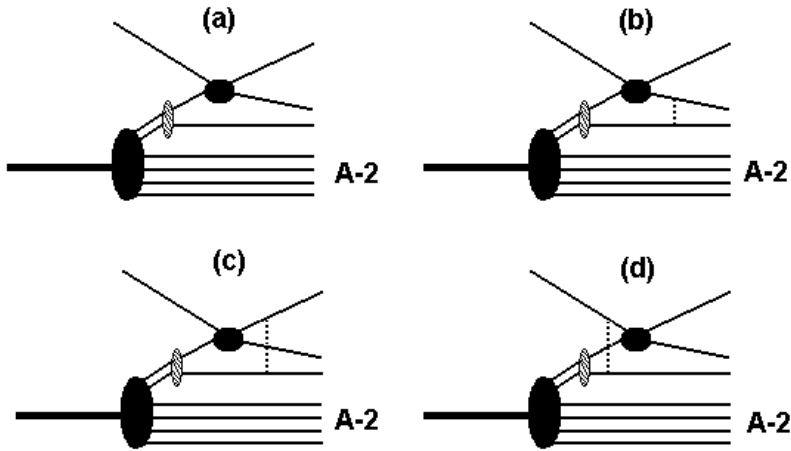}  }
%\end{hayrmer}  
\caption{Diagrams describing PWIA~(a), final~(b,c), and initial~(d) state 
reinteractions for two-nucleon correlations}  
\label{Fig.4}  
\end{figure}

The effect of the rescatterings in the SRC (in the range of 
$\alpha<0.8$ or $\alpha >1.2$) 
can be accounted for by  introducing a correction factor $\kappa$ 
which  multiplies the SRC spectral function of Eq.(\ref{sfunrest}).
We define  $\kappa$ as follows:
\begin{equation}
\kappa = {|F_a + F_b + F_c + F_d|^2\over |F_a|^2},
\label{kappa}
\end{equation}
where $F_a$ is the  PWIA  amplitude, and 
$F_b$, $F_c$ and $F_d$ are the single rescattering amplitudes corresponding 
to   $p+ (NN)_{SRC} \to p + N + N$ scattering shown in  Figure 4. 
To obtain the $F$'s we use the rescattering amplitudes for the $d(p,pp)n$  
reaction calculated in Ref.\cite{FPSS97}: 
\begin{equation}
F_{(j)} = -{(2\pi)^{3\over 2}\over 4 i}A^{hard}_{pp}(s,t)\int 
{d^2k_t\over (2\pi)^2}f^{pN}(k_t)(\psi^{\mu}_d(\tilde p^{(j)}_s) - 
n\cdot i \psi^{\prime \mu}(\tilde p^{(j)}_s)),
\label{F_j}
\end{equation} 
where j(n)=b(1),c(1),d(-1).
$A^{hard}_{pp}$ is the amplitude of the $pp$ hard scattering which, 
within the factorization approximation, cancels in $\kappa$.
$f^{pN}$ is the amplitude of a small angle (soft) $pN$ scattering, 
$N$ can be either proton or neutron.
$\psi_d$ is the deuteron wave function and $\psi^\prime$ accounts for 
the distortion due to FSI (see Ref.\cite{FPSS97}).

For higher order rescatterings  we applied the probabilistic approximation of 
Ref.\cite{Yael} which we used already in the case of small Fermi momenta.  
This is justified since  in the 
kinematics of two-nucleon SRC the second and higher order rescatterings 
happen outside of the SRC. It is worth noting that the error originating 
from the last approximation is rather small since for 
the intermediate size nuclei ($A\sim 12-16$)  the   overall 
contribution of higher order  rescatterings in the considered kinematics 
of the $A(p,2p)X$ reaction is small (a few percent as compared with the single 
rescattering contribution \cite{Yael}).

\medskip

It is important to emphasize that the major qualitative feature of 
reinteractions with uncorrelated nucleons, in high energies, is the existence 
of the approximate conservation law for the light cone momenta of interacting 
particles\cite{FSS,MS}. Namely, for energetic particles small angle soft 
reinteractions do not change the $\alpha$- distribution. 

To demonstrate this let us consider the propagation of fast nucleon with 
momentum $p_1= (E_1,p^z_{1}, 0)$ through the nuclear medium.  
After the small angle 
reinteraction  of this  nucleon with a nucleon in the nucleus 
with momentum $p_2=(E_2,p^z_{2},p^t_{2})$, the energetic 
nucleon still maintains its high momentum and leading  $z$-direction having 
now a momentum $p^{\prime}_1=(E^{\prime}_1,p^{z\prime}_{1},p^{t\prime}_{1})$ 
with ${<(p^{t\prime}_{1})^2>\over (p^{z\prime})^2}\ll 1$.  The other nucleon momentum 
after the collision is $p^{\prime}_2=(E^{\prime}_2,p^{z\prime}_{2},p^{t\prime}_{2})$. 
The energy momentum  conservation for 
this scattering allows us to write for the ``$\alpha$'' component:
\begin{equation}
\alpha_1 + \alpha_2 = \alpha^\prime_1 + \alpha^\prime_2 \equiv
{p_{1-}\over m} + {p_{2-}\over m}  = {p^\prime_{1-}\over m} +
{p^\prime _{2-}\over m}.
\label{claw}
\end{equation}
The change of the $\alpha_2$  (``$-$'')
component due to rescattering can be obtained from Eq.(\ref{claw}):
\begin{equation}
\Delta \alpha_2\equiv {\Delta p_{2-}\over m} =  {p_{2-}-p^\prime_{2-}\over m}
= {p^\prime_{1-}-p_{1-}\over m}\ll 1.
\label{dal}
\end{equation}
which means:
\begin{equation}
{\alpha^\prime_2 \approx \alpha_2.}
\label{dal1}
\end{equation}
In Eq.(\ref{dal}) we use the conditions  
${p^\prime_{1-}\over m}, {p_{1-}\over m} \ll 1$ 
which is well satisfied  in the small angle reinteractions since 
${<(p^{t\prime}_{1})^2>\over (p^{z\prime})^2}\ll 1$.
Thus, with the increase of the incident energy a 
new approximate conservation law is emerging:
$\alpha_2$ is conserved by ISI/FSI. 
The uniqueness of the high energy rescattering is in the fact that although both 
the energy and the momentum of the nucleons are distorted by the  
rescattering, the combination of $E_2-p^z_2$ is almost not affected.
In the same way the rescattering of the incoming and 
two outgoing protons  in the (p,2p) reaction 
conserve the reconstructed $\alpha$-component of the target proton.
Therefore the $\alpha$-distribution measured in $A(p,2p)X$ reaction 
reflects well the original $\alpha$-distribution of the target proton  in
the nucleus. A numerical estimate of this conservation will be presented 
in the next section.

\medskip

To complete the discussion on ISI/FSI we 
should mention that for incident proton momenta exceeding $6-9~GeV/c$
the Glauber approximation overestimates the 
absorption of protons if compared with the data of Ref.\cite{Carroll,EVA}. 
The overestimate of the absorption in these experiments is attributed to the 
Color Transparency (CT) phenomena, in which it is assumed that the 
hard $pp\to pp$ primary process in the $A(p,2p)X$ reaction is 
dominated by the interaction of protons in the point like $qqq$ configurations.
As a result, immediately before and after the hard interaction the color 
neutral PLC has a diminished strength for ISI/FSI reinteraction.  
Since the PLC is not an eigenstate of QCD Hamiltonian 
(free nucleons have a finite size) the interaction strength will evolve to 
the normal hadronic interaction strength in parallel with the evolution of 
PLC to the normal hadronic size during the propagation of the fast proton in 
the nuclear medium. We estimate the CT phenomenon within the quantum diffusion 
model of Ref.\cite{FFLS}. This model which  describes reasonably 
well\cite{FSZ93} the data \cite{Carroll} assumes the following amplitude 
for the $PLC-N$ soft interaction:
\begin{equation}
f^{PLC,N}(z,k_t,Q^2) = i\sigma_{tot}(z,Q^{2}) \cdot 
e^{{b\over 2 }t}\cdot {G_{N}(t\cdot\sigma_{tot}(z,Q^{2})/\sigma_{tot})
\over G_{N}(t)},  
\label{F_NNCT}  
\end{equation}  
where $b/2$ is the slope of elastic $NN$ amplitude,    $G_{N}(t)$ 
($\approx  (1-t/0.71)^{2}$) is the Sachs form factor and 
$t= -k_t^2 $. The last factor in Eq.(\ref{F_NNCT}) accounts for
the difference  between elastic  scattering of  PLC and average
configurations, using  the observation that the $t$-dependence
of $d\sigma^{h+N\to h+N}/dt $ is roughly that of 
$\sim~G_{h}^{2}(t)\cdot G_{N}^{2}(t)$ and that $G_{h}^{2}(t)\approx 
exp(R_h^2t/3)$, where $R_h$ is the rms radius of the hadron.

In Eq.~(\ref{F_NNCT}) $\sigma_{tot}(l,Q^{2})$  is the  effective total 
cross section for the PLC to interact  at distance $l$ from 
the hard interaction point and $\sigma_{tot}$ is the pN total cross section. 
The quantum diffusion model~\cite{FFLS} predicts:
\begin{equation}
\sigma _{tot}(,Q^{2}) = \sigma_{tot} \left \{ \left ({z \over l_{h}} + 
{\langle r_{t}(Q^2)^{2} \rangle \over \langle r_t^{2}  \rangle } 
(1-{z \over l_{h}}) \right )\Theta (l_{h}-z) + \Theta (z-l_{h})\right\}, 
\label{SIGMA_CT}  
\end{equation} 
where ${l_h = 2p_{f}/\Delta~M^{2}}$, with ${\Delta~M^{2}=0.7-1.1~GeV^{2}}$. 
Here ${\langle r_{t}(Q^2)^{2} \rangle}$  is the average squared transverse 
size of the  configuration  produced at the interaction  point.  
In several realistic models considered  in Ref.\cite{FMS92} it can be 
approximated as ${ {\langle r_{t}(Q^2)^2\rangle\over\langle r_t^2\rangle} 
\sim{1\,GeV^2\over Q^2}}$ for  $Q^2~\geq~1.5~GeV^2$.  Note that due to 
expansion, the results of the calculations are rather insensitive 
to the value of this ratio whenever it is much less than unity.
For numerical calculations we assumed $\Delta M^2\approx 0.7 GeV^2$  
as was  chosen to describe  the nuclear transparencies from
$A(p,2p)X$ \cite{Carroll} and 
$A(e,e'p)X$ \cite{NE18}  experiments (see comparisons in Ref.\cite{FSZ93}).

\begin{figure}[hb]  
\centerline{  
\epsfig{width=4.2in,file=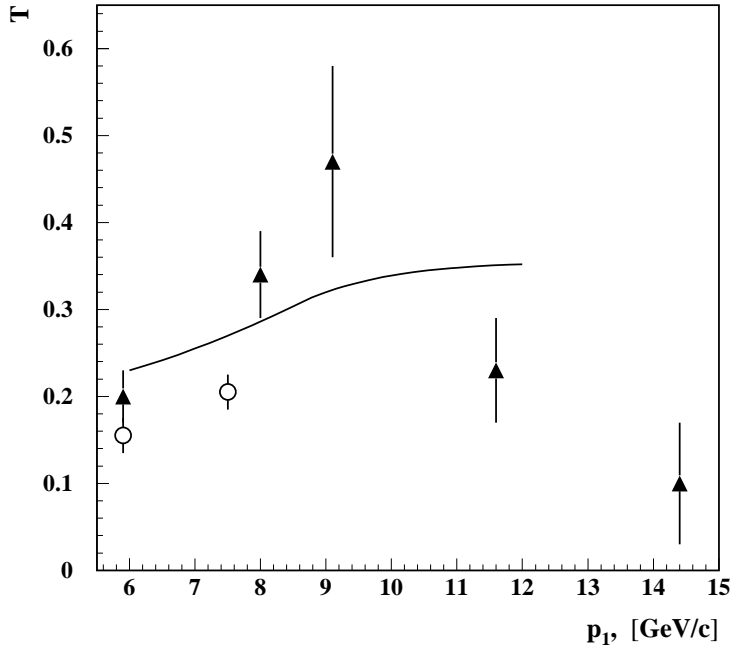}  }
\caption{The $p_1$-dependence of the transparency T calculated 
within quantum diffusion model. Data marked by triangles and 
circles are from [32] and [36] respectively.}  
\label{Fig.5}  
\end{figure}  

In Figure 5 we compares the prediction of quantum diffusion model for nuclear 
transparency $T$ with the data of the EVA experiment\cite{kn:I101,EVA}. 
The transparency $T$ is defined as the ratio of the $A(p,2p)X$ cross section 
calculated using PWIA, color screening and rescattering effects to the cross 
section calculated within PWIA only.  The comparison shows that one has a fair 
agreement with the data up to 9 GeV/c incoming proton momenta (note that one 
expects that the probabilistic model of rescattering to work within $20$\% accuracy). 
The decrease of the experimental values of transparency can be understood 
in terms of the interplay of the hard and soft component in the 
amplitude of high momentum transfer pp scattering\cite{RP,BT} which is not 
incorporated  in the current calculations.
Since in the further analysis we will 
concentrate only in the region of incoming proton momenta $5.9\le p_1\le 7.5  GeV/c$ 
where this interplay does not play a role, we will use the simple formulae of 
Eqs.(\ref{F_NNCT},\ref{SIGMA_CT}) for numerical  estimations. The detailed analysis of 
the energy dependence of the nuclear transparency, $T$ will be  presented elsewhere.

\begin{figure}[hb]  
\centerline{  
\epsfig{width=4.2in,file=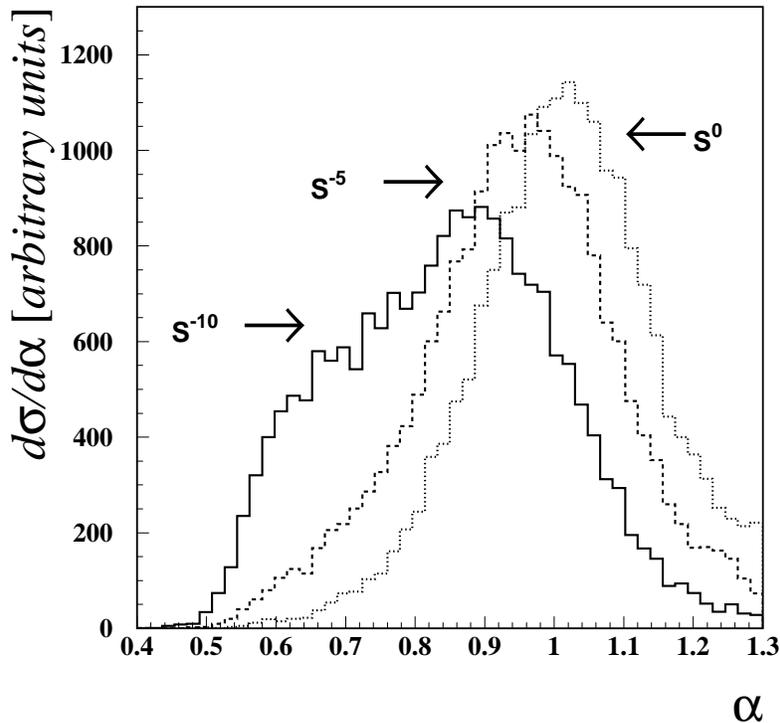}  }
\caption{The $\alpha$-dependence of the $A(p,2p)X$ cross section for different
assumed  $s$-dependences  of the hard elastic $pp$ scattering cross section.}  
\label{Fig.6}  
\end{figure}  

\section{Results of the Model}
\label{III}

In the following chapter we discuss the results of the model 
presented in Chapter II, for several nuclear observables that can be 
measured in  the $A(p,2p)X$  reaction. 
We are particularly interested in two kinds of information: how the
substructure of high-momentum transfer $pp$ scattering reveals itself
in the nuclear reaction and what kind of information 
one can infer about short-range nuclear structure from these reactions.
For numerical calculations in this chapter we apply the kinematics of 
EVA experiment\cite{kn:I101}. Because of the multidimensional character of 
the kinematical restrictions the numerical  calculations are implemented 
through the Monte Carlo calculation. Furthermore, we will present the cross 
sections in arbitrary units  since we are interested mainly in the shapes of 
the $\alpha$ and $p_t$ dependence of the $A(p,2p)X$ cross section.

\subsection{How the Quark Substructure of Hard pp Scattering
is being Reflected in the Nuclear Observables}
\label{IIIa}

The power law energy-dependence of the hard elastic  $pp$ scattering
cross section is the signature of  the dominance of quark-gluon degrees 
of freedom in the high-momentum  transfer  scattering (see e.g. \cite{hex}).
As was predicted in Ref.\cite{FLFS} if this strong energy-dependence  
($\sim s^{-10}$) exists in the nuclear medium it will amplify the 
contribution to the cross section coming from the scattering off deeply 
bound protons. These protons have a large momentum in  
the direction of the incoming proton. 

\begin{figure}[hb]  
\centerline{  
\epsfig{width=4.2in,file=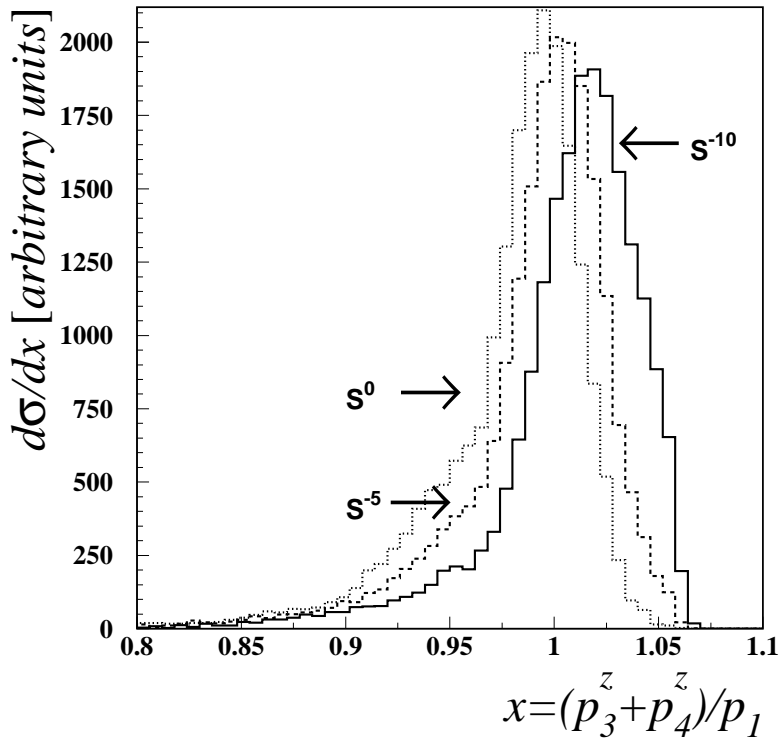}  }
\caption{The $x $-dependence of the cross section for different
hard elastic $pp$ scattering power laws.}  
\label{Fig.7}  
\end{figure} 

Since the cross section for the high-momentum transfer scattering of 
incoming proton off the bound proton at fixed and large 
$\theta_{cm}\sim 90^0$ is roughly proportional to $(\alpha s)^{-10}$
(see Eqs.(\ref{pwia}-\ref{pp})),  an  observation that reflects the 
sensitivity of $A(p,2p)X$ reaction to the high momentum component of 
the nuclear wave function is the shift of the $\alpha$-spectra 
to the lower $\alpha$ values. To
demonstrate this sensitivity, in Figure 6 we represent the $\alpha$-dependence 
of the  $A(p,2p)X$ reaction cross section assuming different $s$-dependences 
of the cross section for hard $p+p\to p+p$ scattering. 
These calculations are merely for illustration of the connection between 
the s-dependence and the  $\alpha$-shift.
Figure 6 confirms that the larger is the negative power of $s$-dependence 
for the  hard $pp$ scattering  the larger is the average longitudinal momentum
of the interacting bound nucleon ($\alpha <1$).

The $\alpha$-shift also produces an excess
of the total longitudinal momentum of the final outgoing protons as
compared to the initial longitudinal momentum  $p_1$.  One can
characterize this excess through the variable:
\begin{equation}
x = {p^z_3 + p^z_4\over p_1},
\label{x}
\end{equation}
which will increase as the power of the hard $pp$ scattering
cross section increases. In Figure 7 we show the calculated $x$-dependence of 
the cross section for different assumed s-dependences.
The expected shift to the higher x (lower $\alpha$) 
is clearly seen in Figure 7. The x-distribution for
quasielastic $C(p,2p)X$ reactions peaks at $x<1$, if one assumes 
no s-dependence of the  elementary $p+p\to p+p$ reaction. 
As the dependence on s increases the peak is shifted to $x>1$ which represents 
the nuclear ``boosting'' effect: the outgoing protons have more 
longitudinal momentum than the incoming momentum. It is worth noting that 
this effect is reminiscent of subthreshold production in nuclei, in which 
a very low available energy in the nuclear medium can cause 
dramatic changes in the cross section  of the reaction.

\begin{figure}[hb]    
\centerline{  
\epsfig{width=4.2in,file=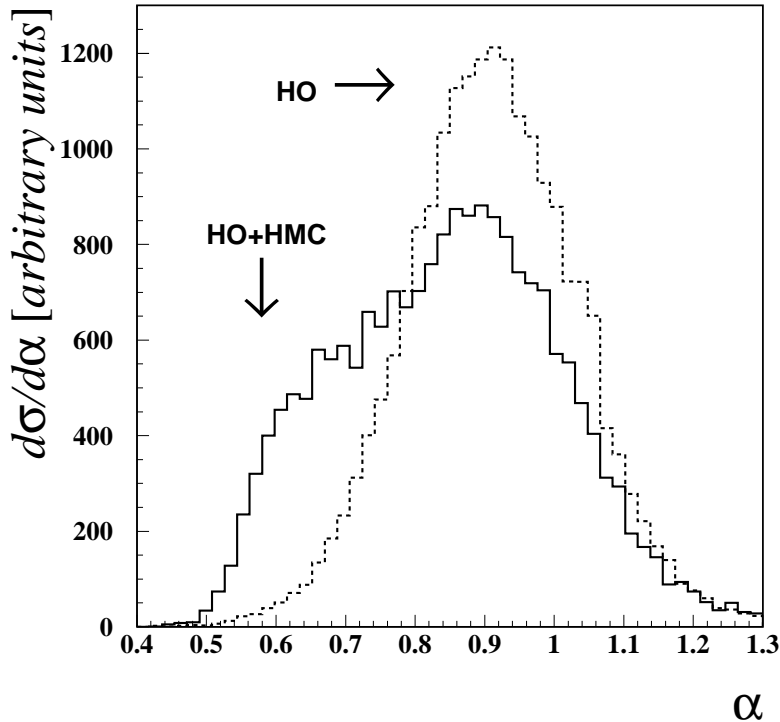}  }
\caption{The $\alpha-$dependence of the cross section calculated for two 
models of the nuclear wave functions. ``HO'' is Harmonic Oscillator and 
``HO+HMC'' - corresponds to the 
short range correlation model of Section \ref{IIc}. The 
$A(p,2p)X$ cross section is calculated within PWIA at $p_1=6~GeV/c$ and 
$\theta_{cm}=90^0$.}  
\label{Fig.8}  
\end{figure}

\subsection{Sensitivity to Short Range Correlations in  Nuclei}
\label{IIIb}

The next question we would like to address is the sensitivity  of the
$\alpha$-shift to the existence of high momentum components in the 
nuclear ground 
state wave function. To asses this sensitivity we compare the cross 
sections of the $A(p,2p)X$  reaction using  two models for the nuclear wave 
function: an Harmonic Oscillator  (HO) model and the two-nucleon SRC 
model of high momentum component (HMC) of nuclear wave function, described in  
Section \ref{IIc} (HO+HMC).  In Figure 8 we present the $\alpha$-dependence 
of the 
$A(p,2p)X$ cross section calculated within PWIA at $p_1=6~GeV/c$ and 
$\theta_{cm}=90^0$ using these two models.

As Figure 8  shows, even at moderate energies as $p_1 = 6~GeV/c$ the $\alpha$- 
dependence shows substantial sensitivity to the high momentum structure of 
the nuclear wave function. Thus, the measured cross section at 
small $\alpha$ will allow us to obtain the characteristics of the high 
momentum tail of the wave function. 

In Figure 9, we show the results of the PWIA calculations for transverse 
momentum distribution of the cross section of $A(p,2p)X$ reaction. It also 
exhibits a sensitivity to the high momentum part of the nuclear 
wave function. However, as will be shown below, unlike the  
$\alpha$-distribution the  transverse momentum distribution 
is strongly distorted due to the initial and final state 
interactions. Note that hereafter, for the transverse missing momentum 
distribution, we will consider only the $p_y$ component of $p_t$. This 
restriction is related to the fact that the experimental data 
have better resolution for the $p_y$ component of missing momentum.

\begin{figure}[hb]  
\centerline{  
\epsfig{width=4.2in,file=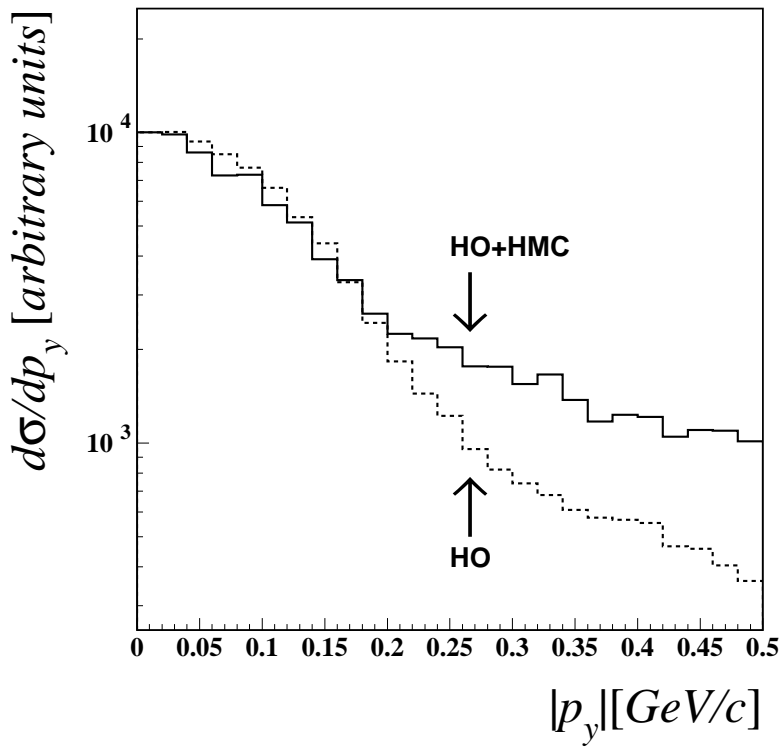}  }
\caption{The $p_{y}$-dependence of the cross section for the two
models of nuclear wave functions described in the text. 
The kinematics of the calculations and notations are the same as in  
Fig.~8.}  
\label{Fig.9}  
\end{figure}

\subsection{The Effect of Initial and Final State interactions}
\label{IIIc}

As was discussed in Chapter 2  (see Eq.(\ref{dal})) one expects that the 
soft rescatterings with uncorrelated nucleons at high energies will conserve 
the $\alpha$ parameter of interacting nucleons. Thus the measured 
$\alpha_2$-distribution of $A(p,2p)X$ cross section will not be 
affected strongly by the ISI/FSI and will  reflect the original 
$\alpha$-distribution of the target proton in the nucleus.

In Figure 10 we compare the $p_2$ and $\alpha$ distribution of 
the $\theta_{cm}=90^0$ $A(p,2p)X$ differential cross section at $p_1=6~GeV/c$. 
The dashed lines correspond to the PWIA prediction, thus representing the 
``true'' momentum distribution of the bound nucleon.
The solid lines  represent the calculation including ISI/FSI. In the latter case 
the $p_2$ and $\alpha$ are reconstructed through 
the momenta of the incoming ($p_1$) and outgoing protons ($p_3$, $p_4$), thus 
representing the ``measured'' quantities.

Notice the effect of the ISI/FSI on the $p_2$-distribution versus the effect 
of the same ISI/FSI on the $\alpha$-distribution. 
As we mentioned before, both the reconstructed energy and momentum of the 
target proton are modified by the rescattering, but their linear combination, 
$\alpha$, is almost unchanged.

\begin{figure}[hb]  
\centerline{  
\epsfig{width=4.2in,file=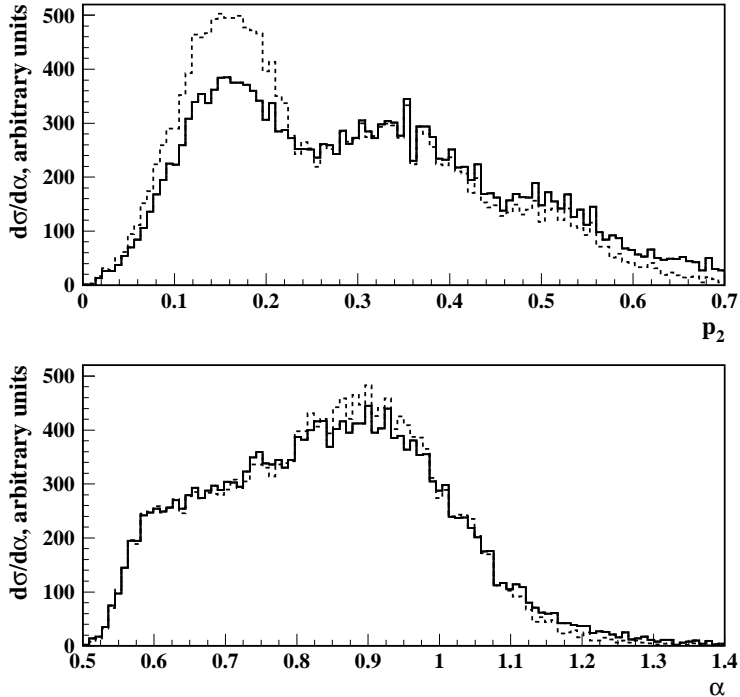}  }
\caption{The $\alpha$-dependence of the cross section 
  with and without rescattering with uncorrelated nucleons.}  
\label{Fig.10}  
\end{figure}  

Finally, in Figure 11  we show the transverse momentum distribution 
($p_{y}$) calculated for the same kinematics as in 
Figure 10. Figure 11 shows substantial ISI/FSI effects on the 
$p_{y}$-distribution for both calculation with and without Color Transparency. 
The large contribution from ISI/FSI in the transverse momentum distribution 
is attributed to the structure of small angle hadronic interaction in high 
energies. The rescattering is mainly transverse thus affecting 
maximally the transverse momenta of interacting nucleons.
  
The above discussion allows us to conclude that the experimental study of 
the $\alpha$-distribution provides direct information on high 
momentum components of the nuclear wave function. On the other hand, the large 
values of missing transverse momentum is mainly sensitive to the 
dynamics of initial and final state interaction. In the subsequent sections 
we will discuss the analysis of the first experimental data on $A(p,2p)X$ 
reaction. 

\begin{figure}[hb]  
\centerline{  
\epsfig{width=4.2in,file=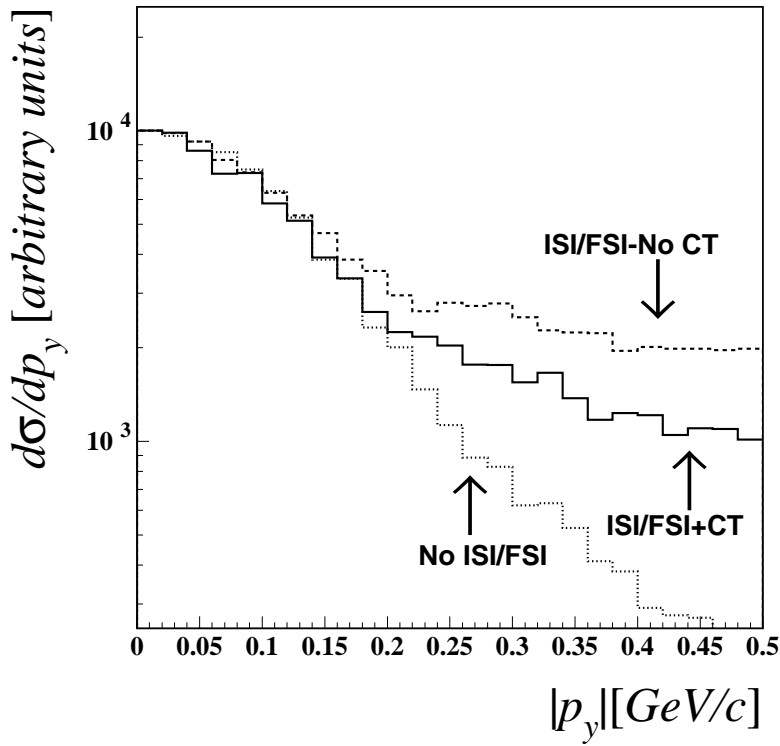}  }
\caption{The $p_{y}$ dependence of the cross section with and without 
rescattering effects.}  
\label{Fig.11}  
\end{figure}  
  
\section{ Measurements and data.}
\label{IV}

We compare the calculations with the data that were collected in EXP 850
using the EVA spectrometer at the AGS accelerator of Brookhaven National
Laboratory. During the preparation of this work these data were the only 
ones on high momentum transfer quasi-elastic reactions \cite{kn:I101}. 
In this chapter we will briefly describe the experiment and the experimental
procedures relevant for comparing the data with the calculations.
In the following  chapter we will present the calculations and 
compare them with the data.

The EVA collaboration performed a second measurement over a wider 
kinematical range  with incident momenta above 7.5 GeV/c. These data 
were not analyzed yet. Some of the calculations in this work are  
predictions for these new data which might become later available.

\subsection{The Experimental Setup}
\label{IVa}

 The EVA spectrometer, located on the secondary line C1,  consisted of 
a 2 meter diameter and 3 meter long super-conducting solenoidal magnet 
operated at 0.8 Tesla (see Fig 12). The beam entered along the $z$ axis and 
hit a series of targets located at various $z$ positions. 
The scattered particles were tracked by four cylindrical chambers
(C1-C4 , Fig 12). Each had 4 layers of long straw drift tubes with a high 
resistance central wire.
For any of the 5632 tubes that fired, the drift time to its central wire 
was read out.
In three out of the four cylindrical chambers signals were  read out at 
both ends, providing position information along the $z$ direction as well.
The straw tubes information allowed the target identification, the 
measurement of the particles transverse momentum 
as they were bent in the axial  magnetic field, and their scattering 
angles.  The overall resolution caused by the beam, the target and the 
detector were determined from the two body elastic pp scattering measurement. 
The standard deviation ($\sigma$) for the resolution of the 
transverse momentum is $\Delta p_t/p_t=7\% $ and 0.27 GeV for the missing
energy. The  polar angles ($\theta_3$,$\theta_4$)
of the two outgoing  protons were measured with a resolution of 7 mrad.
The beams ranged in intensity
from 1 to $2\cdot 10^7$ over a one second spill every 
3 seconds. Two counter hodoscopes in the beam (only one shown in fig 12) 
provided beam alignment and a timing reference and two differential
Cerenkov counters (not shown in fig 12) identified the incident particles. 
Three levels of triggering were used to select events with a 
predetermined minimum transverse momentum. 
The first two  hardware triggers selected events with
transverse momenta p$_t>$ 0.8 and p$_t>$ 0.9 GeV/c, for the 6 and 7.5 GeV/c
measurements, respectively. The third level software trigger required two 
almost coplanar tracks, each satisfying the second level trigger requirement 
and low multiplicity hits in the straw tubes. See Ref \cite{kn:ref7} for a
detailed description of the trigger system. Details on the EVA spectrometer 
are given in Refs. \cite{kn:ref7,kn:ref3,kn:ref4,kn:ref5}.

  Three solid targets, CH$_2$, C and CD$_2$ (enriched to $95\%$) were
placed on the $z$ axis inside the C1 cylinder separated by about 20 cm. 
They were $5.1$x$5.1$  $cm^2$ squares and 6.6 $cm$ long in the $z$ 
direction except 
for the CD$_2$ target which was 4.9 cm long. Their positions  were
interchanged at several intervals in order to reduce systematic uncertainties
and to maximize the acceptance range for each target. Only the C target
was used to extract the QE events, while the other targets served for
normalizations and references.

\subsection{Event Selection and Kinematical Constraints}

Quasi-elastic scattering events, with only two charged particles in the
spectrometer, were selected. An excitation energy of the residual nucleus
$\mid E_{miss}\mid$$<$ 500 MeV was imposed in order to suppress events
where additional particles could be produced without being detected in
EVA. 
Since this cut is above $m_\pi$, some inelastic background,   
such as those coming from $pA\to pp \pi^0 (A-1)$ events,  
could penetrate the cuts and had to be subtracted.  
The shape of this background was determined
from a fit to  the $E_{miss}$ distribution of  events with extra tracks 
in the spectrometer. An inelastic background with this shape was subtracted. 
The measured distributions represent background subtracted quantities.
See Refs. \cite{kn:ref5,kn:I101} for more details.

 \begin{figure}[htbp]
  \begin{center}
    \leavevmode
\epsfig{angle=-90,width=4.2in,file=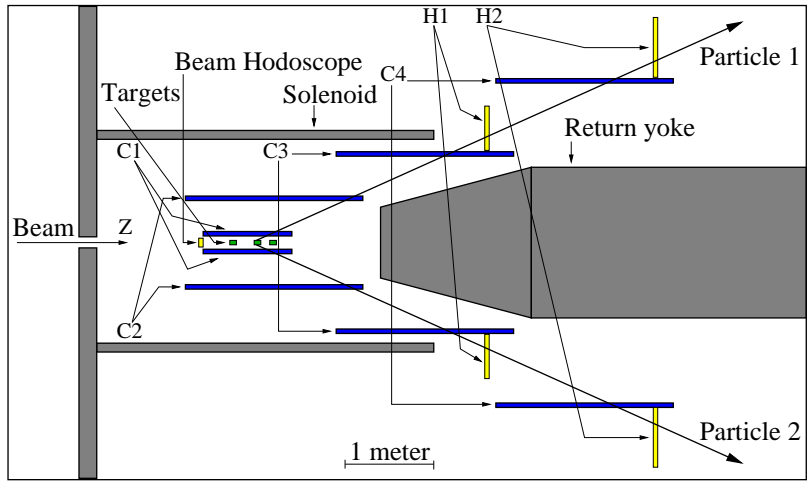} 
\vspace{0.5cm}
   \caption{ A schematic view of the  EVA spectrometer. C1-C4 are the 
straw tube 4 layers detectors. H1-H2 are scintillator hodoscopes used 
for fast triggering on high $p_t$ events. The three targets in C1 are
shown in typical positions. The beam direction (symmetry axis of the detector)
is chosen to be the z axis. Not shown in the figure are the beam counters
upstream the spectrometer as well as the full iron structure around the 
solenoid.} 
    \label{Fig.12}
  \end{center}
\end{figure}

The coordinate system was chosen with the $z$ coordinate in the beam 
direction and the $y$ direction  normal to the scattering plane ($x,z$). 
The latter is defined by the incident beam and  one of the emerging
protons. The selection among the two was random. This arbitrariness in
the selection does not affect the extracted quantities of interest.
The data were analyzed in terms of the momenta in the $y$ direction
$p_{y}$ and the light cone $\alpha$ variable. 

 $\alpha$ was determined with a precision of $\sigma\simeq 3\%$. 
The $ p_{y}$ (perpendicular to the scattering plane)  had a
resolution of $\sigma= 40$ MeV/c. The resolution in $p_{x}$ (in the 
scattering plane)  was $\sigma=170$ MeV/c. Because of its better 
resolution, $p_{y}$ was used to represent a transverse component. 

 The laboratory polar angles of both detected  protons were
limited by a software cut to a region of $\pm(3-5)^o$ around the 
center of the angular acceptance, for each target position. The 
angular range enforced by the software cut is smaller than the geometrical
limits of the spectrometer (see Fig 12) but it ensures a uniform acceptance.
Since the experiment was focused on shapes and not absolute values,
an acceptance correction in the $(\theta_3, \theta_4)$ plane
is not needed. 
An explicit cut on the center of mass scattering angle 
$\theta_{cm}$ was not applied on the data,
however the cuts on the laboratory polar angles limit 
the $\theta_{cm}$ to the range of 83$^o$ to 90$^o$ for the proton at rest 
kinematics.

\subsection{The Longitudinal ($\alpha$) Distributions}
\label{IVc}

 Each target position corresponds to a limited polar angular range 
$(\theta_3, \theta_4)$ and $\alpha$ is a strong function of 
$\theta_3 + \theta_4$. To cover the largest possible acceptance 
in $\alpha$ one has to merge the measured  $\alpha$-distributions
from different targets.
The distributions from the individual target positions were normalized to 
each other using the overlapping regions. The experimental error
in each bin includes also the relative normalization error. The 
value of $\mid \theta_3 - \theta_4 \mid$ was limited by the largest 
common acceptance of all target position.

To summarize:  the following angular acceptance cuts were applied 
on the data:

\begin{itemize}

\item

 $\mid \theta_3 - \theta_4 \mid<0.06$ radians (For all target positions and 
both beam energies).         

\item
 downstream target: $ 23.5^0  <\theta_3<32.0^0$ and 
$23.5^0 < \theta_4 < 29.5^0$ or $ \theta_3$ and  $\theta_4$ inverted. 

\item
middle target: $ 20.0^0 <\theta_3<30.0^0$ and
$22.0^0 < \theta_4 < 28.0^0$ or $ \theta_3$ and  $\theta_4$ inverted.
 
\item
upstream target: $ 19.0^0 <\theta_3<28.0^0$ and
$21.0^0 < \theta_4 < 27.5^0$ or $ \theta_3$ and  $\theta_4$ inverted.

\end{itemize}
 
 These cuts yield for 5.9 GeV/c the following $\alpha$ acceptance ranges:

\begin{itemize}

\item
downstream target:  $0.9<\alpha<1.05$.
\item
middle  target:     $0.767<\alpha<0.967$.
\item
upstream target:    $0.7<\alpha<0.867$.

\end{itemize}

For the 7.5 GeV/c data the angular ranges were:

\begin{itemize}

\item
 downstream target: $ 22.0^0  <\theta_3<32.0^0$ and 
$22.0^0 < \theta_4 < 31.5^0$ or $ \theta_3$ and  $\theta_4$ replaced.

\item
middle target: $ 21.0^0 <\theta_3<27.0^0$ and
$21.0^0 < \theta_4 < 27.0^0$. 
 
\item
upstream target: $ 20.0^0 <\theta_3<26.0^0$ and
$20.0^0 < \theta_4 < 26.0^0$.

\end{itemize}
 
These cuts yield for 7.5 GeV/c the following $\alpha$ acceptance ranges:

\begin{itemize}

\item
downstream target:  $0.967<\alpha<1.05$.
\item
middle  target:     $0.834<\alpha<1.0$.
\item
upstream target:    $0.767<\alpha<0.934$.

\end{itemize}

\subsection{The Transverse ($p_{y}$) Distributions}
\label{IVd}

 The $p_{y}$-distributions were studied for narrow regions of $\alpha$.
The regions of $\alpha$ were chosen to yield a large overlap between the 
5.9 GeV/c and the 7.5 GeV/c data sets for each target position:

\begin{itemize}

\item
$0.74<\alpha<0.84$      for the upstream target position.

\item
$0.82<\alpha<0.92$       for the middle  target position.

\item
$0.95<\alpha<1.05$.     for the downstream target position.

\end{itemize}

 The shape of the $p_{y}$- distributions for the two data at
6 and 7.5 GeV/c are consistent in each one of the three $\alpha$- regions. Since
the  data sets of the two energies were found to be consistent they were 
added in order to reduce the statistical errors. Even after this procedure 
the poor statistics for the $0.95<\alpha<1.05$ range do not allow us to draw 
conclusions for this range. All the data presented consist of events that 
passed all the quasi-elastic cuts and the residual inelastic background 
was subtracted  in a way similar to that described for the 
$\alpha$- distributions (see Ref. \cite{kn:I101,kn:ref5} for details). 
All measured $p_{y}$- distributions are normalized to 10000 at $p_{y}=0$ and 
shown on a logarithmic scale to emphasize their shapes. The data 
are compared to the calculations in chapter 5.

\section{Comparison of the calculations with the data}
\label{V}

\subsection{The Longitudinal ($\alpha$) Distributions}
\label{Va}

As  was mentioned in Chapter 3 the calculations are implemented through the 
Monte Carlo code which allowed to incorporate the theoretical calculations 
with the multidimensional kinematic cuts applied in the experiment.
The following cuts have been included in the calculations:
\begin{itemize}
\item The angular and $\alpha$ acceptances are constrained 
       for  the same ranges as presented in chapter \ref{IV} for the data.
\item
   $60^{0}<\theta_{cm}<120^{0}$ (for all target positions).
\end{itemize}
The calculations include all considered nuclear effects (EMC, ISI/FSI and CT).

Figure 13 shows the measured longitudinal $\alpha$-distributions at
5.9 GeV/c and 7.5 GeV/c together with the calculations. 
In the  calculation  we used the two-nucleon correlation model 
for the  high momentum component of the nuclear wave function, discussed 
in Chapter \ref{II}. For the parameter $a_2(^{12}C)$ which defines the 
strength of the SRC in the nuclear spectral function 
(Eq.(\ref{sfunrest})) 
we used the estimate obtained from the analysis of high $Q^2$ and large 
Bjorken x  $A(e,e')X$ data Ref.\cite{DFSS}. This analysis  yield  
$a_2\approx 5$ for $^{12}C$.

The $\chi^2$ per degree of freedom obtained by comparing the measured and 
calculated distributions at 5.9 GeV/c and 7.5 GeV/c ($\chi^2=0.8$ and 
$\chi^2=2.0$ respectively) confirm that the calculation and the data have 
the same shape. 

The next question we ask is whether the data allow us to understand the 
ingredients contributing to the strength of the $\alpha$-distribution 
at lower $\alpha$-values.

First we check if the 
high momentum transfer elastic $pp$ scattering  off bound nucleon still
attains  the $s^{-10}$ energy dependence. 
In Figure 14 we  compare the calculations done using $s$-independent pp 
cross section (triangle points) and the $pp$ cross section parameterized 
according to Eq.(\ref{pp}), in which $\frac{d\sigma^{pp}}{dt} \propto s^{-10}$ 
(solid points). 
If there was no scaling for hard $pp$ scattering  in the nuclei 
the $\alpha$-distribution would  peak around $\alpha=1$, as shown by the 
calculations with no "s-weighting" (triangles).
The data clearly show a shift to lower $\alpha$ which confirms
the strong s-dependence of the quasi elastic process.

%-------------------------------------------------------------------------
 \begin{figure}[H]
  \begin{center}
    \leavevmode
    \centerline{\epsfig{file=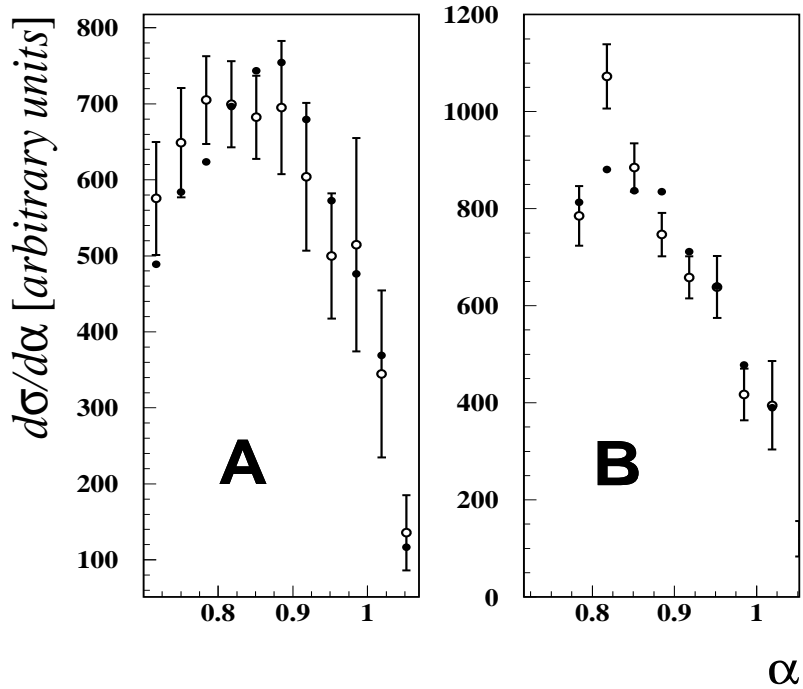,width=4.4in,height=10cm}}
   \caption{A comparison between calculated $\alpha$-distributions
     ($\bullet$) and the experimental data ($\circ$) at 5.9
     GeV/c (A) and 7.5 GeV/c (B).} 
    \label{Fig.13}
  \end{center}
\end{figure}
%-------------------------------------------------------------------------

Next  we address the question whether the strength seen at $\alpha<1$ 
comes from the SRC in nucleus. Figure 15 shows two calculated 
$\alpha$-distributions for the incoming proton momentum 
of 5.9 GeV/c. 
One distribution is calculated with the harmonic oscillator wave
function only (i.e. $a_2=0$, in Eq.(\ref{sfunrest})) (Triangle points). 
The  second distribution is calculated with the SRC contribution to the  
high momentum tail of the nuclear wave function, described by
$a_2=5$ (solid points).  The  open circles are the data.
It is clearly seen in the figure that the $\alpha$-distribution
calculated with $a_2=0$ does not provide sufficient strength at 
low $\alpha$ to describe the data, and SRC contributions are necessary.

It is important to note that both the strong $s$-dependence of  
hard $pp$ scattering and the contribution of SRC are needed 
for agreement with the data. A mean field wave function 
for the nucleus would require a very unreasonable (exponentially falling 
with s) energy dependence of the $pp$ scattering cross section, in order 
to explain the observed strength of the cross section at $\alpha<1$. 
Moreover the agreement with the data using 
the same value of $a_2$ parameter obtained from electronuclear reactions  
indicates that we are dealing with a genuine property of the nucleus 
that does not depend on a specific probe.

%-------------------------------------------------------------------------
 \begin{figure}[H]
  \begin{center}
    \leavevmode
    \centerline{\epsfig{file=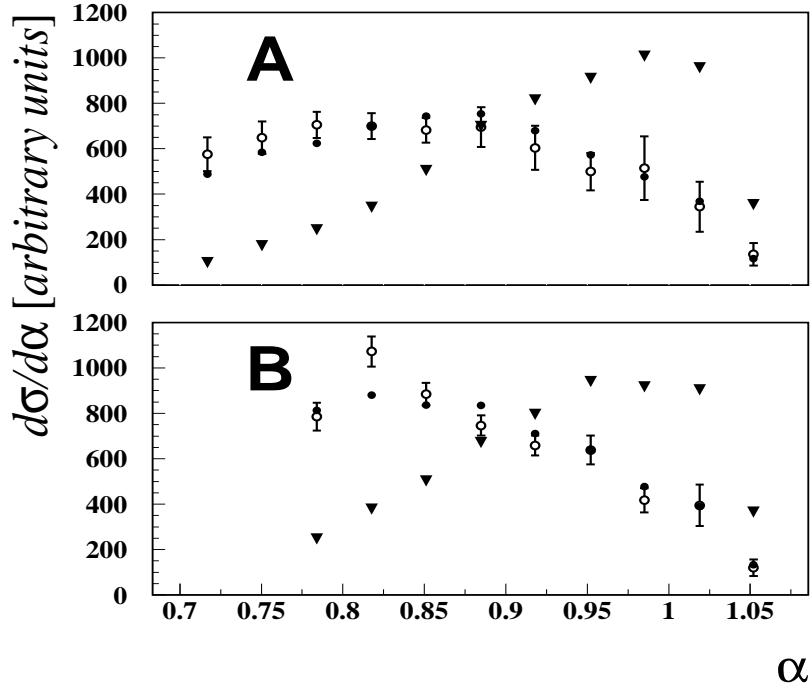,width=4.4in,height=10cm}}
     \caption{Calculated longitudinal $\alpha$-distributions with
     ($\bullet$) and without ($\nabla$) "s-weighting" compared to the
     measured data ($\circ$), at 5.9 GeV/c (A) and 7.5 GeV/c (B).}
    \label{Fig.14}
  \end{center}
\end{figure}
%-------------------------------------------------------------------------

%-------------------------------------------------------------------------
 \begin{figure}[H]
  \begin{center}
    \leavevmode
    \centerline{\epsfig{file=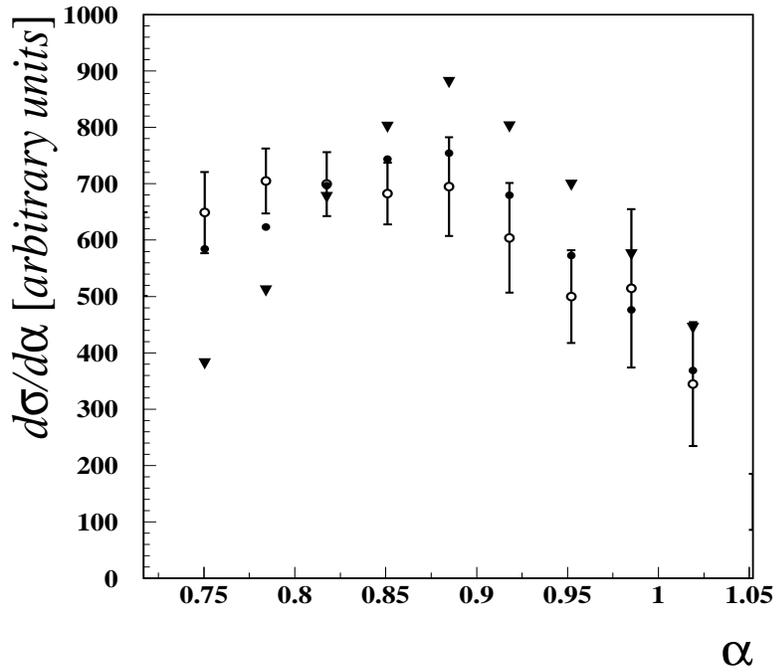,width=4.4in,height=10cm}}
   \caption{Longitudinal $\alpha$-distributions for 5.9
     GeV/c.($\circ$ - data, $\nabla$ - calculations with $a_2=0$,
     $\bullet$ - calculations with $a_2=5.0$).}
    \label{Fig.15}
  \end{center}
\end{figure}
%-------------------------------------------------------------------------

\subsection{The Transverse ($p_{y}$) Distributions}
\label{Vb}

As it was discussed in  Sections \ref{II}  and \ref{III}, we expect  
the transverse missing momentum of the quasielastic $A(p,2p)X$ cross section 
to be sensitive mainly to the dynamics of ISI/FSI. The studies of 
electro-nuclear $A(e,e'p)X$ reactions, 
in which FSI occurs through the rescattering of only one knocked-out proton 
demonstrated that the eikonal 
approximation can describe the FSI with better than 10\% 
accuracy (see e.g. \cite{Garrow}). This indicates that the expected level of 
accuracy in calculations of ISI/FSI in $A(p,2p)X$ reactions, in which one 
incoming and two outgoing protons undergo the soft rescatterings, will be on 
the order of 15-20\%. Keeping these accuracies in mind we compare the 
theoretical calculations with the data checking how well the probabilistic 
approximation of ISI/FSI can reproduce the shape of the transverse missing 
momentum distribution.

%-------------------------------------------------------------------------
 \begin{figure}[H]
  \begin{center}
    \leavevmode
    \centerline{\epsfig{file=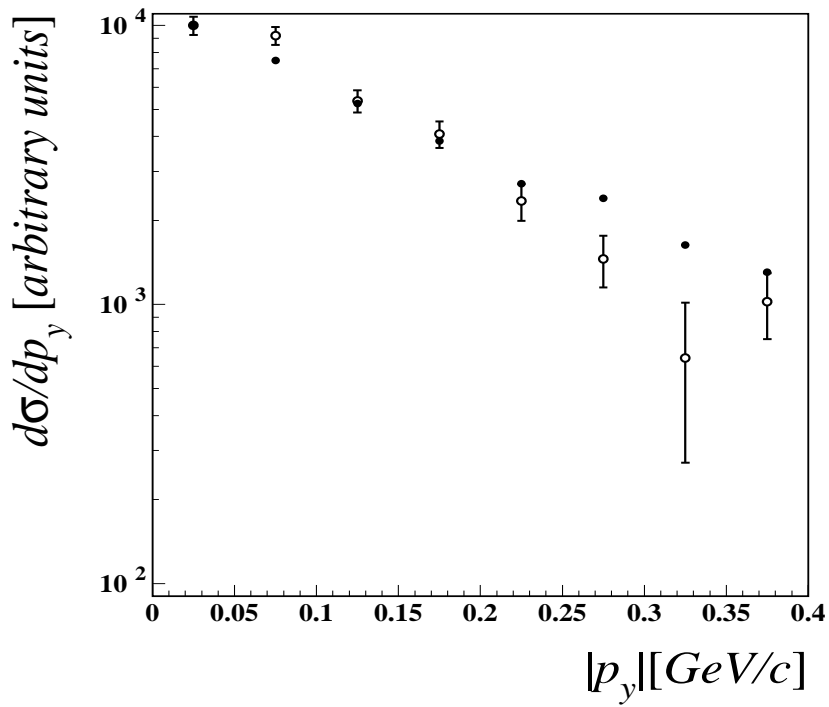,width=4.4in,height=10cm}}
   \caption{A comparison between the calculated ($\bullet$) and  
    experimental ($\circ$) $p_{y}$-distribution
     combined for 5.9 GeV/c and 7.5 GeV/c momenta. The kinematics   
     for the upstream target with $0.74<\alpha<0.84$ is used 
     (see text for details).}
    \label{Fig.16}
  \end{center}
\end{figure}
%------------------------------------------------------------------------

The following kinematical constraints are imposed in the Monte Carlo
calculations
\begin{itemize}
\item
   middle target:  $0.82<\alpha<0.92$.
\item
   upstream target:  $0.74<\alpha<0.84$.
\item
   $\mid\theta_3-\theta_4\mid<0.06$ rad (for all target positions).
\item
   $60^{0}<\theta_{cm}<120^{0}$ (for all target positions).
\end{itemize}
The calculations include all the effects discussed in the Chapters 2 and 3
(i.e.  ISI/FSI, EMC, CT) and the strength of the SRC defined with 
$a_2=5$.

Figure 16 shows the comparison between the measured transverse $p_{y}$
distribution and the calculated distribution. 
The theoretical and experimental distributions are normalized to 1000 at 
the first bin so only the difference in shape between them is relevant.
They are for the combined 5.9 GeV/c and 7.5 GeV/c data and the upstream target
($\alpha = 0.79 \pm 0.05$). See chapter 4 for the detailed procedure
of combining the 5.9 GeV/c and 7.5 GeV/c data sets. We followed the same 
procedure in the calculations. Figure 17 shows the similar to Figure 16 comparison 
for the kinematics of the middle stream target 
($\alpha=0.87 \pm 0.05$). 

The calculations presented in Figure 16 and 17 
overestimate the data at the transverse missing momenta above $0.2 ~GeV/c$. 
There are several reasons for such a discrepancy. 
First, one should notice that the tail of the distribution above   
$p_{y}=200 MeV/c$ is only $10\%$ of the peak value at $p_{y}=0$.
Since calculation and the data are normalized at the maximum, even small 
discrepancy between calculation and the data at $p_{y}=0$ will reproduce 
a large discrepancy at large values of $p_y$.

Next, this discrepancy may be the indication of  the limit of applicability of 
the probabilistic approximation of ISI/FSI. In this approximation we 
neglected  the  interference terms which may contribute at large values of 
transverse momenta. Indeed as the complete 
calculation of $d(p,2p)n$ reaction demonstrated\cite{FPSS97} the interference terms 
are not negligible at $p_t\ge 150-200 MeV/c$ and their contribution  tends to 
diminish the overall cross section.

%-------------------------------------------------------------------------
 \begin{figure}[H]
  \begin{center}
    \leavevmode
    \centerline{\epsfig{file=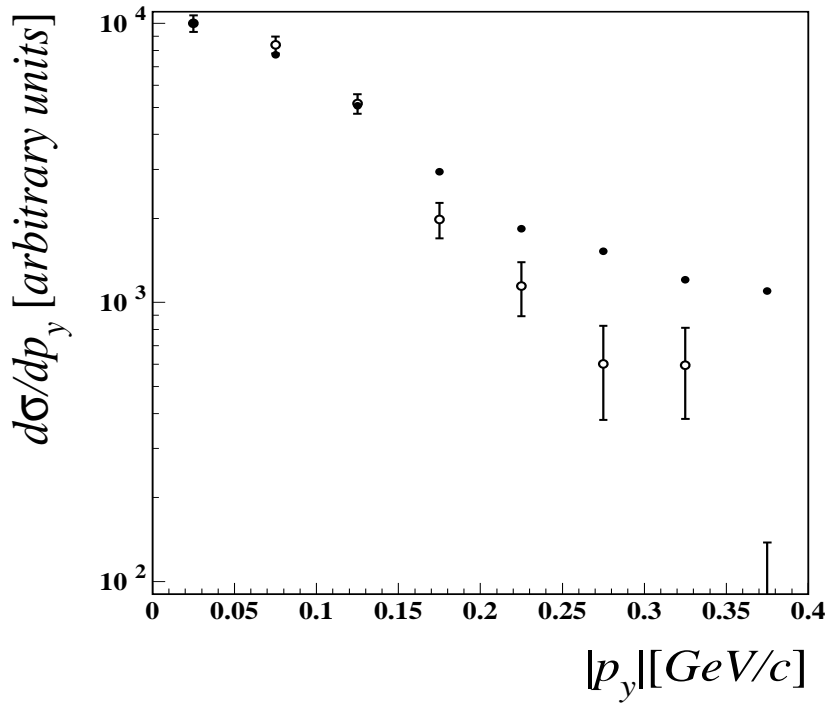,width=4.4in,height=10cm}}
   \caption{A comparison of calculated ($\bullet$) and
    experimental ($\circ$) $p_{y}$ distributions for combined
     5.9 GeV/c and 7.5 GeV/c energies. The kinematics of the middle stream 
     target with $0.82<\alpha<0.92$ is used (see text for details).}
    \label{Fig.17}
  \end{center}
\end{figure}
%------------------------------------------------------------------------

Another reason for  the discrepancy may be the fact that within 
the eikonal approximation, starting at transverse missing momenta 
($\geq 150-200 MeV/c$) the ISI and FSI are dominated by incoherent elastic 
rescattering which enhance the cross section of the nuclear reaction 
(see for the details Ref.\cite{Yennie}). It was observed 
in Refs.\cite{EFGMSS,FGMSS} that incoherent elastic rescatterings are 
much more sensitive to the CT phenomena then the nuclear absorption is. 
The qualitative reason is that the absorption is proportional to the 
total cross section of PLC-N interaction, $\sigma^{tot}_{PLC,N}$, 
while incoherent elastic rescattering is proportional to  
$(\sigma^{tot}_{PLC,N})^2$. Thus the overestimate of the calculation 
may indicate that the  onset of CT is stronger than it is modeled in 
the calculations (see Section \ref{II}). Note that a noticeable 
($\sim 20\%$)  
change in the strength of the incoherent elastic rescattering will result 
only  $\sim 5\%$ change of the absorption thus such a 
modification of the size of the CT effect will still maintain the agreement 
of the calculation with the transparency data of Ref.\cite{Carroll}.

Ending the above discussion we can only conclude that the  strength of 
the high transverse momentum distributions is generated by ISI/FSI. However 
both improved  theoretical calculation of ISI/FSI and the better experimental 
resolution are needed for understanding the details of the
dynamics behind the strength of high  transverse momentum distributions.

\section{ Summary}

We present the theoretical analysis of the first published data on the 
high momentum transfer quasielastic $C(p,2p)X$ reaction.

First, we outline the light cone plane wave impulse approximation, in which 
the high momentum component of the nuclear wave function is treated within 
a two-nucleon short range correlation model. Within the same model  
it was predicted in Ref.\cite{FLFS} that the $\alpha$-distribution 
of the  $A(p,2p)X$ cross section will be shifted to the smaller 
values of $\alpha$ thereby enhancing the contribution from SRC.
We further develop the SRC model taking into account the medium 
modification of the bound nucleon  as well as initial and final state 
reinteractions of the incoming and two outgoing protons in the nuclear medium,
combined  with the color transparency effects.

For nuclear medium modification we demonstrated that within the color screening 
model, which describes  reasonably the available electroproduction data, 
the strength of the SRC is not obscured.
Furthermore we demonstrated that in the high energy regime the 
$\alpha$-distribution of the bound proton is 
practically unaltered by ISI/FSI.  As a result the 
$\alpha$-distribution of the  $C(p,2p)X$ cross section reflects the 
genuine distribution of the bound proton in the nucleus. We also 
showed that the transverse missing momentum distribution is strongly 
sensitive to the dynamics of initial and final state reinteractions, 
and discussed its potential use to study  the effects related to the color 
transparency phenomena.

In addition  to the $\alpha$ and $p_t$ distributions we discussed 
the dependence of the cross section on the 
total longitudinal momentum of the two outgoing protons.  It 
indicates the existence of a nuclear ``boosting'' effect, in which 
the longitudinal momentum of the two outgoing protons is larger than 
the  momentum of the incoming proton. This result is in qualitative 
agreement with the new data recently obtained at EVA\cite{EVA}. 

After briefly describing the experiment we proceed with comparison of
the theoretical calculations with the data.  The comparison demonstrates
that the  theoretical  expectation of the $\alpha$ shift, based on scaling 
in hard elastic scattering off a bound nucleon in the nucleus, is correct.
The physical meaning of these shifts is  that
hard quasi elastic $pp$ scattering is sensitive to the
high momentum components of the nuclear wave function.
One observes that a  momentum tail in the nuclear wave function 
that is needed to explain the data  is significantly larger than what is 
expected from the mean field approximation. The value of the two nucleon SRC
strength  needed to describe the data is in agreement with the SRC
strength obtained from electronuclear reactions.
The analysis of the transverse missing momentum distribution shows that
it is very sensitive to the mechanism of ISI/FSI  and  both improved calculations 
and the data are needed for understanding the  details of the dynamics that generates 
the high transverse momentum strength. Thus the studies of the transverse-momentum 
distribution may emerge as an additional tool for study the color transparency
phenomena.

\section*{Acknowledgements}

Part of the data related to $p_y$ distribution  have not been published before.
We would like to acknowledge the EVA  collaboration allowing us to present them 
in this paper. The authors are thankful to the EVA collaboration, especially to 
the spokespersons:~S.~Heppelmann and A.~Carroll for very useful discussions. 
Special thanks to Y.~Mardor for providing the details of her analysis of 
the experimental data. 

M.~Sargsian gratefully acknowledges a contract from Jefferson Lab under which this 
work was done. The Thomas Jefferson National Accelerator Facility (Jefferson Lab) 
is operated by the Southeastern Universities Research Association (SURA) under 
DOE contract DE-AC05-84ER40150. 
This work is supported also by DOE grants under contract
DE-FG02-01ER-41172 and  DE-FG02-93ER-40771 as well as by the U.S. - Israel 
Binational Science foundation and the Israel Science Foundation founded by 
the Israel Academy of Sciences and Humanities.


\references

\bibitem{hex}S.J.~Brodsky and G.R.~Farrar, Phys. Rev. Lett. {\bf 31}, 1153
        (1973); Phys. Rev. {\bf D11}, 1309 (1975);  
        V.~Matveev, R.M.~Muradyan and A.N.~Tavkhelidze, Lett. Nuovo
        Cimento {\bf 7}, 719 (1973).
\bibitem{Isgur_Smith}N.~Isgur and C.H.~Llewellyn Smith, Phys. Rev. Lett.
        {\bf 52}, (1984) 1080;  Phys.Lett. {\bf B217},   535 (1989).
\bibitem{Rady}A.~Radyushkin, Acta Phys. Pol. {\bf B15}, 403 (1984).
\bibitem{BCL79} S.~J.~Brodsky, C.~E.~Carlson and H.~J.~Lipkin,
        Phys.\ Rev.\  {\bf D20}, 2278 (1979).
\bibitem{FGST79} G.~R.~Farrar, S.~Gottlieb, D.~Sivers and G.~H.~Thomas,
        Phys.\ Rev.\  {\bf D20}, 202 (1979).
\bibitem{RS95}G.~P.~Ramsey and D.~Sivers, Phys.\ Rev.\  {\bf D52}, 116 (1995).
\bibitem{LLP}P.~Landshoff, Phys. Rev. {\bf D10}, 1024 (1974); 
        P.~Landshoff and D.~Pritchard, Z.Phys. {\bf C6}, 69 (1980).
\bibitem{BoSt} J.~Botts and G.~Sterman, Nucl.\ Phys.\  {\bf B325}, 62 (1989).
\bibitem{BoSo}C.~Bourrely and J.~Soffer, Phys.\ Rev.\  {\bf D35}, 145 (1987).
\bibitem{FS88}L.L.~Frankfurt and M.I.~Strikman, Phys. Rep.  {\bf 160}, 235 (1988).
\bibitem{FLFS}G.~R.~Farrar, H.~Liu, L.~L.~Frankfurt, and M.~I.~Strikman
             Phys. \ Rev. \ Lett. {\bf 62}, 1095 (1989).
\bibitem{kn:I101} Y.Mardor {\em et al.}, Phys. Lett. B437( 1998) 257.
\bibitem{Feynman}R.~Feynman, {\em Photon - Hadron Interactions}, 
                 W.A. Benjamin Inc. 1972.
\bibitem{FS81}L.L.~Frankfurt and M.I.~Strikman, Phys. Rep. {\bf 76}, 214 (1981).
\bibitem{DFSS}L.~L.~Frankfurt, M.~I.~Strikman, D.~B.~Day and M.~Sargsian,
              Phys.\ Rev.\  {\bf C48}, 2451 (1993).
\bibitem{CSFS}C.~Ciofi degli Atti, S.~Simula, L.~L.~Frankfurt 
               and M.~I.~Strikman, Phys.\ Rev.\  {\bf C44}, 7 (1991).
\bibitem{BT}S.~J.~Brodsky and G.~F.~de Teramond, Phys.\ Rev.\ Lett.\ 
            {\bf 60}, 1924 (1988).
\bibitem{SBB}D.~Sivers, S.J.~Brodsky and R.~Blankenbecler, 
             Phys. Rep. {\bf 23}, 1 (1976).
\bibitem{FPSS95}L.~Frankfurt, E.~Piasetsky, M.~Sargsian and M.~Strikman,
              Phys.\ Rev.\  {\bf C51}, 890 (1995).
\bibitem{EMC}J.J.~Aubert {\em et al.} (EM Collaboration), Phys. Lett. B {\bf 123}, 
             275 (1983).
\bibitem{FS85}L.~L.~Frankfurt and M.~I.~Strikman, Nucl.\ Phys.\  {\bf B250} 
              (1985) 143.
\bibitem{Frank}M.~R.~Frank, B.~K.~Jennings and G.~A.~Miller, 
               Phys.\ Rev.\ C {\bf 54}, 920 (1996).
\bibitem{FSZ93}L.~L.~Frankfurt, M.~I.~Strikman and M.~B.~Zhalov,
             Phys.\ Rev.\  {\bf C50}, 2189 (1994).
\bibitem{rescaling}R.L. Jaffe, F.E. Close, R.G. Roberts and G.G. Ross: Phys. 
                  Lett. {\bf B134}, 449 (1984).
\bibitem{MSS} W.~Melnitchouk, M.~Sargsian and M.~I.~Strikman,  
              Z.\ Phys.\ A {\bf 359}, 99 (1997).
\bibitem{FMS92}L.~Frankfurt, G.~A.~Miller and M.~Strikman,
               Phys.\ Rev.\ Lett.\  {\bf 68}, 17 (1992).   
\bibitem{Yael} I.~Mardor, Y.~Mardor, E.~Piasetzky, J.~Alster and M.~M.~Sargsian,
                             Phys.\ Rev.\ C {\bf 46}, 761 (1992) 
\bibitem{FSS} L.~L.~Frankfurt, M.~M.~Sargsian and M.~I.~Strikman, 
                           Phys.\ Rev.\ C {\bf 56}, 1124 (1997).
\bibitem{FPSS97} L.~L.~Frankfurt, E.~Piasetzky, M.~M.~Sargsian and M.~I.~Strikman,
                                  Phys.\ Rev.\ C {\bf 56}, 2752 (1997).
\bibitem{MS}M.~M.~Sargsian, Int.\ J.\ Mod.\ Phys.\ E {\bf 10}, 405 (2001).
\bibitem{Carroll} A.~S.~Carroll {\it et al.}, Phys.\ Rev.\ Lett.\  {\bf 61}, 1698 (1988).
\bibitem{EVA} A.~Leksanov {\it et al.},  Phys.\ Rev.\ Lett.\  {\bf 87}, 21230 (2001).
\bibitem{FFLS} G.~R.~Farrar, H.~Liu, L.~L.~Frankfurt and M.~I.~Strikman,
                             Phys.\ Rev.\ Lett.\  {\bf 61}, 686 (1988).
\bibitem{FMS92} L.~Frankfurt, G.~A.~Miller and M.~Strikman,
                                 Comments Nucl.\ Part.\ Phys.\  {\bf 21}, 1 (1992).
\bibitem{NE18}N.~Makins {\it et al.},{\em (NE18 collaboration)} 
              Phys.\ Rev.\ Lett.\  {\bf 72}, 1986 (1994).
\bibitem{Yael2}Y.~Mardor {\em et al.}  Phys.\ Rev.\ Lett.\  {\bf 81}, 5085 (1998).
\bibitem{RP} J.~P.~Ralston and B.~Pire, Phys.\ Rev.\ Lett.\  {\bf 61}, 1823 (1988).
\bibitem{JM} B.~K.~Jennings and G.~A.~Miller, Phys.\ Lett.\ B {\bf 318}, 7 (1993).

\bibitem{kn:ref7} {  J.Y.Wu {\em et al.} , Nuclear Instruments and
    Methods A 349 (1994) 183.}
\bibitem{kn:ref3}M.A.Shupe {\em et al}. {\em EVA, a solenoidal detector
                 for large angle exclusive reactions: Phase I - determining color 
                 transparency to 22 GeV/c.} Experiment E850 Proposal to Brookhaven 
                 National Laboratory, 1988 (unpublished).
\bibitem{kn:ref4}{\em Measurement of the dependence of the $C(p,2p)$
                 cross section on the transverse component of the spectral momentum}, 
                 S.Durrant, PhD thesis, Pennsylvania State University, 1994 (unpublished).
\bibitem{kn:ref5}{\em Quasi-Elastic Hadronic Scattering at Large Momentum
                 Transfer}, Y.Mardor, PhD thesis, Tel Aviv University, 1997 (unpublished).
\bibitem{Garrow}K.~Garrow {\it et al.}, arXiv:hep-ex/0109027, (2001).
\bibitem{Yennie}D.R.~ Yennie, {\em in Hadronic Interactions of Electrons 
                and Photons}, edited by J.~Cummings and D.~Osborn 
                (Academic, New York, 1971), p.321.
\bibitem{EFGMSS}K.Sh.~Egiyan, L.L.~Frankfurt, W.~Greenberg, G.A.~Miller, 
                M.M.~Sargsian and M.I.~Strikman, 
                Nucl.\ Phys.\ A {\bf 580}, 365 (1994).
\bibitem{FGMSS}L.L.~Frankfurt, G.A.~Miller, W.~Greenberg, M.M.~Sargsian and 
                M.I.~Strikman, Z.\ Phys.\ A {\bf 352}, 97 (1995).
\end{document}